\documentclass[aps,pra,twocolumn,amsmath,amssymb,nofootinbib,superscriptaddress]{revtex4-2}
\usepackage{todonotes}
\usepackage{times}
\usepackage{graphics}
\usepackage{dcolumn}
\usepackage{bm}
\usepackage{float} 
\usepackage{amsthm}
\usepackage{braket}
\usepackage{indentfirst} 
\usepackage{xr}  
\usepackage[colorlinks]{hyperref}
\usepackage{xcolor}
\usepackage{paralist}
\usepackage{tabularx}
\usepackage{multirow}
\usepackage{booktabs}
\usepackage{mathtools}   
\usepackage{bm} 
\usepackage{algorithm}
\usepackage{algpseudocode} 

\usepackage[normalem]{ulem}   
\usepackage{array}
\usepackage{longtable}
\usepackage{ragged2e} 

\renewcommand\vec{\mathbf}  
\usepackage{verbatim}
\usepackage{makecell}

\newtheorem{theorem}{Theorem}

\begin{document}
\title{Efficient Measurement Error Mitigation with Subsystem-Balanced Pauli Twirling}
\author{Xiao-Yue Xu} 
\affiliation{Henan Key Laboratory of Quantum Information and Cryptography, Zhengzhou, Henan 450000, China}

\author{Chen Ding}
\affiliation{Henan Key Laboratory of Quantum Information and Cryptography, Zhengzhou, Henan 450000, China}
 
\author{Wan-Su Bao}
\email{bws@qiclab.cn}
\affiliation{Henan Key Laboratory of Quantum Information and Cryptography, Zhengzhou, Henan 450000, China}

\date{\today}
\begin{abstract} 
  Measurement error mitigation (MEM) is essential for realizing reliable quantum computation. Model-free measurement error mitigation (MF-MEM) is an important class of MEM methods that employs Pauli twirling—typically with a random twirling set—to convert measurement noise into a state-independent scaling factor, thereby enabling error mitigation through simple calibration. However, such methods face prohibitive sampling overhead, limiting their scalability. To address this, we introduce subsystem-balanced Pauli twirling (SB-PT), a twirling method designed for MF-MEM that enforces Pauli operator balance on measuring subsystems to selectively suppress dominant measurement noise. Theoretically, for a weight-$\tau$ Pauli observable, SB-PT removes all independent error components using only $\mathcal{O}(4^\tau)$ random circuits, substantially reducing the sampling overhead over conventional Pauli twirling. This efficiency gain is most significant for sparse observables. To extend such resource-efficient mitigation to arbitrary observables, we develop a hardware-efficient measurement transformation framework that converts high-weight Pauli operators into low-weight effective ones via linear-depth circuits. The circuit noise introduced during this transformation is jointly mitigated with native measurement noise using a unified twirling protocol, ensuring robust performance. Extensive numerical simulations demonstrate a greater than 16-fold improvement in sampling efficiency over conventional random twirling, with consistent performance gains across varying system sizes and error regimes. This work provides a resource-frugal and experimentally viable path toward high-fidelity measurement in near-term quantum devices.
\end{abstract}
\maketitle      
\section{Introduction} 

Quantum computers hold significant potential for solving classically intractable problems across a wide range of fields, including quantum chemistry, optimization, and machine learning~\cite{Nielsen2002QuantumComputation,Ladd2010,Peter1999Shor}, yet are fundamentally limited by hardware imperfections. Among these, measurement errors pose a particularly critical challenge: they systematically corrupt the final readout after quantum state collapse, irreversibly destroying computational results. Crucially, the dimensionality of measurement noise scales exponentially with qubit number, requiring $\mathcal{O}(4^n)$ parameters for full characterization, rapidly becoming infeasible for practical devices.

This exponential barrier motivates simplified error models. The most direct approach leverages measurement's projective nature, treating readout errors as classical state transitions after state collapse~\cite{Chen2019Detector,Bravyi2021Mitigate,Maciejewski2020mitigationofreadout,Kwon2021Ahybrid,Nation2021Scalable}.
We model them through a stochastic map with transfer matrix $\Lambda$ relating ideal ($p_{\text{ideal}}$) and noisy ($p_{\text{noisy}}$) distributions via 
\begin{equation}\label{eq-lambda}
    p_{\text{ideal}} = \Lambda^{-1}p_{\text{noisy}}. 
\end{equation}
Quantum mechanically, this corresponds to a quantum channel $\mathcal{N} = \mathcal{E}_\Lambda \circ \mathcal{D}$ where the complete dephasing channel $\mathcal{D}(\rho) = \sum_k \langle k|\rho|k\rangle |k\rangle\langle k|$ eliminates coherence, followed by the population noise channel $\mathcal{E}_\Lambda(\sigma) = \sum_i (\sum_j [\Lambda]_{ij}\sigma_j)|i\rangle\langle i|$ for the resulting diagonal state $\sigma=\sum_j\sigma_j|j\rangle\langle j|$ with $\sigma_j=\langle j|\sigma|j\rangle$, which modifies the diagonal elements according to the transfer matrix $\Lambda$.  
However, even this classical reduction remains exponentially costly as it requires $\mathcal{O}(2^n)$ parameters, motivating further simplification. 
The Tensor Product Noise (TPN) model~\cite{Bravyi2021Mitigate,Geller2020Rigorous,Michael2020Efficient,Maciejewski2021modelingmitigation} assumes qubit-independent measurement noise, which implies a transfer matrix that factorizes as $\Lambda_{\text{TPN}} = \bigotimes_{i=1}^n \Lambda_i$. This reduces the complexity to $\mathcal{O}(n)$, making mitigation tractable for large systems, though at the cost of neglecting correlated and quantum coherent errors. 
 
An alternative paradigm is model-free measurement error mitigation (MF-MEM)~\cite{Harrigan2021Quantum,Hicks2021Readoutrebalancing,Alistair2021Qubit,Berg2022Modelfree,Tang2022vna}, which circumvents model-dependent limitations. This method employs Pauli twirling~\cite{Geller2013Efficient,Wallman2016Noisetailoring,Erhard2019Characterizing,Hashim2021Randomizedcompiling} to tailor the full measurement noise channel into a Pauli form, enabling its mitigation via state-independent rescaling without requiring explicit characterization. While effective for moderate-scale systems~\cite{Kim2023Scalable}, the conventional approach using undirected random twirling becomes prohibitively resource-intensive at scale.

Motivated by tailored twirling techniques that construct minimal twirling sets for specific error channels~\cite{Cai2019Constructing}, we introduce Subsystem-Balanced Pauli Twirling (SB-PT) for MF-MEM. SB-PT systematically exploits the structure of measurement noise by designing twirling sets in which each Pauli operator is balanced on the measured subsystems, thereby selectively suppressing the dominant independent error components. We rigorously prove that for any Pauli observable of weight $\tau<n$, SB-PT requires only $\mathcal{O}(4^\tau)$ random circuits to exactly cancel all independent measurement errors, while also delivering strictly tighter sampling overhead bounds than conventional random twirling. To extend these advantages beyond sparse observables, we integrate a measurement transformation protocol that generalizes Compression Readout~\cite{Ding2023Compression,Ding2023quantum}; this protocol converts any $m$-weight Pauli observable into $k$-weight form ($k\leq m$) via $\mathcal{O}(n)$-depth circuits while maintaining compatibility with twirling. This framework incorporates both measurement noise and circuit-induced noise into a unified effective noise model, enabling unified error mitigation. Numerical simulations confirm these advantages across multiple noise regimes and observable types, demonstrating significantly improved resource efficiency over MF-MEM with random twirling while maintaining comparable fidelity. 

The remainder of this paper is organized as follows. Section~\ref{sec2} establishes the theoretical foundation, introducing the Pauli-transfer-matrix representation and the MF-MEM framework. This prepares the essential formalism for our approach. In Section~\ref{sec3}, we develop the SB-PT protocol: we first characterize the algebraic structure of classical measurement noise within the Pauli-transfer-matrix representation, then construct the SB-PT method by exploiting the inherent symmetries. Section~\ref{sec4} addresses mitigation scalability through measurement transformation techniques, enabling efficient SB-PT implementation for high-weight observables. Comprehensive numerical simulations validating our protocol under various noise regimes are presented in Section~\ref{sec5}. Finally, we conclude with discussions on broader implications and potential extensions.

\section{Twirling-based Measurement error mitigation}\label{sec2}
\subsection{The Pauli-Transfer-Matrix formalism}
Measurement noise systematically corrupts the expectation values of Pauli observables by introducing non-physical shifts.  
To establish this formally, we adopt the Pauli-transfer-matrix (PTM) representation~\cite{Endo2018Practical,Greenbaum2015aqm}. Within this formalism, the quantum state $\rho$ is encoded as the column vector $|\rho\rangle\rangle$, while the quantum channel $\mathcal{E}$---completely positive trace-preserving (CPTP)~\cite{Nielsen2002QuantumComputation} map---corresponds to a real matrix $\mathcal{M}_{\mathcal{E}}$ acting on the vector.
The observable $O$ maps to row vector $\langle\langle O|$ in the same basis. 
The expectation value $\langle O \rangle$ of an observable $O$, given that the state $\rho$ has evolved under the channel $\mathcal{E}$, is given by the Hilbert-Schmidt inner product:
\begin{equation}
\langle O \rangle = \text{Tr}(O\mathcal{E}(\rho)) = \langle\langle O| \mathcal{M}_{\mathcal{E}} |\rho\rangle\rangle.
\end{equation}  

We order the $n$-qubit Pauli group $\{I,X,Y,Z\}^{\otimes n}$ by assigning to each index $p\in\{0,1,...,4^n-1\}$ the unique Pauli operator $P_p=\bigotimes_{i=1}^n \sigma_{\vec{p}_i}$, where $\sigma_0=I$, $\sigma_1=X$, $\sigma_2=Y$, $\sigma_3=Z$ and $\vec{p}=(\vec{p}_1,\vec{p}_2,...,\vec{p}_n)\in\mathbb{Z}_4^n$ is the base-4 representation of the number $p$, where each $\vec{p}_i$ is a digit of $p$. Likewise, the subgroup $\{I,Z\}^{\otimes n}$ is ordered by an analogous binary map that assigns to each index $r\in\{0,1,...,2^n-1\}$ the operator $Z_r=\bigotimes_{i=1}^n\sigma_{\vec{r}_i}$, where $\vec{r}\in\mathbb{Z}_2^n$ and $\sigma_0=I$, $\sigma_1=Z$. To embed the subgroup $\{I,Z\}^{\otimes n}$ into the full $n$-qubit Pauli group,
we introduce the index-preserving maps:
\begin{equation}
    \begin{aligned}
        &\phi:\{0,1,\dots,2^{n}-1\}\to\{0,1,\dots,4^{n}-1\}, \\
        &\tilde{\phi}:\mathbb{Z}_{2}^{n}\to\mathbb{Z}_{4}^{n},
    \end{aligned}
\end{equation} 
where we flip every $1$ in $\mathbb{Z}_2^n$ to $3$ in $\mathbb{Z}_4^n$.  
Formally,  
\begin{equation}
    \begin{aligned}
&\tilde{\phi}(\vec{r})=(3\vec{r}_1,\dots,3\vec{r}_n)\in\mathbb{Z}_4^n,\\
&\phi(r)=\sum_{i=1}^{n}3\vec{r}_i\,4^{\,i-1}.
\end{aligned}
\end{equation} 
This construction ensures that $Z_r = P{\phi(r)}$ for all $r$.

The matrix element of $\mathcal{M}_\mathcal{E}$ is defined for indices $p,q\in \{0,1,...,4^n-1\}$ as 
\begin{equation}
[\mathcal{M}_\mathcal{E}]_{pq}=\frac{1}{2^n}\text{Tr}( P_p\mathcal{E}(P_q)).
\end{equation} 
In particular, any Pauli observable $P_q$ can be represented by the standard-basis row vector $\langle\langle P_q|=\mathbf{e}^{\mathsf{T}}_q= (0,\dots,1,\dots,0)^{\mathsf{T}}$ in the PTM representation, where $\mathbf{e}_q$ has a single nonzero entry (1 at index $q$ corresponding to the position of $P_q$ in the Pauli basis). 



\subsection{Model-free measurement error mitigation via Pauli twirling}\label{sec-tbmem}
We now formalize the impact of measurement noise on expectation-value estimation for Pauli observables and introduce the foundational principles of MF-MEM. 
Without loss of generality, we describe our protocol using Pauli-$Z$ observables, as single-qubit Cliffords can map any Pauli observable to this form.
For general measurement noise $\mathcal{R}$ within the PTM representation, the noisy expectation value for Pauli-Z observable $Z_r$ is
\begin{equation}
    \langle \tilde{Z}_r\rangle_\rho =\langle\langle Z_r|\mathcal{R}|\rho\rangle\rangle=2^n\sum_{j=0}^{4^n-1}[\mathcal{R}]_{\phi(r) j}\cdot \varrho_j(\rho),
\end{equation}
where $\varrho_j(\rho)$ is the $j$-th element of $|\rho\rangle\rangle$, which equals the scaled noiseless expectation value $\langle\langle \mathcal{P}_j|\rho\rangle\rangle=\frac{1}{2^n}\text{Tr}(\rho P_j)$. The measurement noise $\mathcal{R}$ results in additional systematic bias of the ideal value $\varrho_{\phi(r)}(\rho)$. 

MF-MEM methods exploits Pauli twirling to convert the measurement noise channel $\mathcal{R}$ into stochastic Pauli channel, enabling bias correction. For a randomly chosen twirling set $\mathcal{S}$ (a subset of the $n$-qubit Pauli group with $|\mathcal{S}| \leq 4^n$), the twirled noise channel is 
\begin{equation}
    \mathcal{R}^{\text{twirled}}=\frac{1}{|\mathcal{S}|}\sum_{P_q\in\mathcal{S}}\mathcal{P}_q \mathcal{R} \mathcal{P}_q,
\end{equation}
where $\mathcal{P}_q$ is the PTM of Pauli operator $P_q$, a diagonal matrix with element $[\mathcal{P}_q]_{ii} = \eta(P_q, P_i)$ for $i\in\{0,1,...,4^n-1\}$. Here $\eta(P_q, P_i) = \pm 1$ denotes the commutator sign satisfying $P_q P_i P_q^\dagger = \eta(P_q, P_i) P_i$. The latter $\mathcal{P}_q$ is virtually applied through post-processing of the measurement outcomes. 
Specifically, when the Pauli term on a single qubit is $X$ or $Y$, this virtual application is achieved by bit-flipping the measurement outcomes (0 and 1).

The PTM elements of the twirled channel simplify to 
\begin{equation}
    [\mathcal{R}^{\text{twirled}}]_{ij}=\frac{1}{|S|}\sum_{P_q\in\mathcal{S}}\eta(P_q,P_i)\eta(P_q,P_j)[\mathcal{R}]_{ij}.
\end{equation}
Defining the scaling factor:
\begin{equation}
    \alpha_{ij}(\mathcal{S})=\frac{1}{|\mathcal{S}|}\sum_{P_q\in\mathcal{S}}\eta(P_q,P_iP_j),
\end{equation}
and noting that $\eta(P_q, P_i) \eta(P_q, P_j) = \eta(P_q, P_i P_j)$ holds under the standard Pauli multiplication convention, we obtain:
\begin{equation}
    [\mathcal{R}^{\text{twirled}}]_{ij}=\alpha_{ij}(\mathcal{S})[\mathcal{R}]_{ij}.
\end{equation}
Crucially, the diagonal elements remain unchanged since:
\begin{equation}
 \alpha_{ii}(\mathcal{S})=\frac{1}{|S|}\sum_{P_q\in\mathcal{S}}\eta(P_q,P_iP_i)=\frac{1}{|S|}\sum_{P_q\in\mathcal{S}} \eta(P_q,I)=1,   
\end{equation}
preserving the intrinsic noise characteristics, while off-diagonal elements are scaled by $\alpha_{ij}(\mathcal{S})$, with the scaling magnitude determined by $\mathcal{S}$. 

When employing the full twirling set $\mathcal{S}=\{I,X,Y,Z\}^{\otimes n}$, the off-diagonal elements vanish ($\alpha_{ij}(\mathcal{S})=0$ for $i\neq j$), resulting in a diagonal matrix $\mathcal{R}^{\text{twirled}}$. Consequently, the twirled noisy expectation value for $Z_r$:
\begin{equation}
    v(\rho)=2^n\sum_{j=0}^{4^n-1}\alpha_{\phi(r) j}(\mathcal{S})[\mathcal{R}]_{\phi(r) j}\cdot \varrho_j(\rho)
\end{equation}
then simplifies to
\begin{equation}
    v(\rho)=2^n[\mathcal{R}]_{\phi(r) \phi(r)}\cdot \varrho_j(\rho)=[\mathcal{R}]_{\phi(r) \phi(r)}\cdot\text{Tr}(\rho Z_r).
\end{equation}
This shows that the noisy expectation value $v(\rho)$ is equal to a scalar multiple of the ideal expectation value $\langle Z_r\rangle\rho=\text{Tr}(\rho Z_r)$. The multiplicative factor $\lambda = [\mathcal{R}]_{\phi(r) \phi(r)}$ depends solely on the measurement noise channel $\mathcal{R}$ and is independent of the state $\rho$.
Since $\text{Tr}(\rho_0 Z_r) = 1$ with $\rho_0=|0\rangle\langle0|^{\otimes n}$ for any $r$ (as $|0\rangle\langle0|$ is a $+1$ eigen-state of $Z$), we obtain $\lambda = v(\rho_0)$. 
The mitigated expectation value is then given by
\begin{equation}\label{eq-tbmem}
a(\rho) = \frac{v(\rho)}{v(\rho_0)}=\frac{\lambda \langle Z_r\rangle_\rho}{\lambda}=\langle Z_r\rangle_\rho,
\end{equation}
which exactly recovers the ideal expectation value.

However, when a reduced twirling set $\mathcal{S}$ (with $|\mathcal{S}| < 4^n$) is employed, the twirling becomes imperfect. This results in non-vanishing off-diagonal elements, meaning $\alpha_{\phi(r) j}(\mathcal{S}) \neq 0$ for some $j \neq \phi(r)$. Consequently, certain noise terms are not fully canceled via Eq.~\ref{eq-tbmem}, introducing error terms of the form $\alpha_{\phi(r) j}(\mathcal{S})[\mathcal{R}]_{\phi(r) j} \cdot \varrho_j(\rho)$ into $v(\rho)$. These errors cause the mitigated estimate $a(\rho)$ to deviate from the ideal expectation value.

We analyze the estimation error of Eq.~\eqref{eq-tbmem}, formally presented in Theorem~\ref{theo:tb-bound} (see proof in Appendix~\ref{sec-proof1}).

\begin{theorem}\label{theo:tb-bound}
For any $n$-qubit quantum state $\rho$ and Pauli-$Z$ observable $Z_r$, with probability at least $1 - \delta$ over the choice of random $\mathcal{S}$, the estimation error $\epsilon(\rho)=|a(\rho)-\text{Tr}(\rho Z_r)|$ is bounded by 
\begin{equation}\label{eq-old_bound}
    \epsilon(\rho)\leq \frac{2\kappa(S)\sum_{j\neq \phi(r)}|[\mathcal{R}]_{\phi(r) j}|}{|[\mathcal{R}]_{\phi(r)\phi(r)}-\kappa(\mathcal{S})\sum_{j\neq\phi(r)}|[\mathcal{R}]_{\phi(r) j}||},
\end{equation}
where
\begin{equation}
    \kappa(\mathcal{S})=\sqrt{\frac{2}{|\mathcal{S}|}(2n\ln 2+\ln(2/\delta))},
\end{equation}
under the diagonal dominance condition:
\begin{equation}
    |[\mathcal{R}]_{\phi(r) \phi(r)}|>\kappa(\mathcal{S})\sum_{j\neq \phi(r)}|[\mathcal{R}]_{\phi(r) j}|.
\end{equation}
\end{theorem}
The above theorem provide a performance guarantee for the MF-MEM methods with a reduced twirling set.

\begin{figure}[t]
    \centering
    {\includegraphics[width=0.48\textwidth]{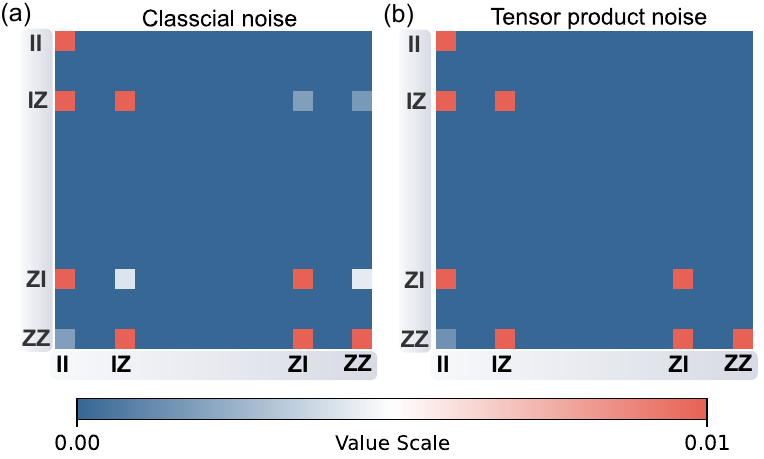}} 
    \caption{\textbf{Pauli Transfer Matrix structure for classical measurement noise models.} 
    To reveal the non-zero patterns more clearly, the heatmaps display the absolute values of the PTM elements. This removes visual clutter caused by the signs of the values. Explicit operator annotations on axes show the correspondence between $n$-qubit Pauli operators and index positions. 
    (a) The PTM of classical measurement noise for two-qubit system. Non-zero elements are confined to the Pauli-$Z$ subset $\mathbb{I}_Z$, with axes annotated by their corresponding Pauli-$Z$ operators. 
    (b) The PTM of tensor product noise (TPN) for two-qubit system. Non-zero elements satisfy the trigger condition $\text{supp}(Z_r) \subseteq \text{supp}(Z_s)$, forming the TPN subset $\mathbb{T}$. Here we define the support of $Z_r$ as $\text{supp}(Z_r) = \{ k \mid \vec{r}_k = 1 \}$.
    The reduced structure results from spatial measurement noise independence.
    Both PTMs rely on calibration data from \textit{Zuchongzhi} 2.1 taken directly from two-qubit measurements and assembled from tensor products of single-qubit data, respectively.} 
    \label{fig0}
\end{figure}

\section{Subsystem-balanced pauli twirling}\label{sec3}
\subsection{Structure of classical measurement noise model}
The efficacy of any tailored twirling strategy hinges on a precise understanding of the noise it aims to suppress. Unlike general quantum processes, classical measurement noise exhibits a highly constrained algebraic structure. This structure arises because the measurement process projects the quantum state into a classical outcome, confining noise effects to a specific subset of elements in the PTM. Furthermore, as evidenced by experimental characterizations of superconducting quantum processors~\cite{Arute2019Quantum, Wu2021Strong, Morvan2024Phase}, this noise is not unstructured; it predominantly follows a hierarchical model where local error processes dominate. This characteristic structure motivates the design of twirling sets that target and cancel these dominant errors efficiently, rather than treating the noise as a black box.

\begin{figure*}[!htbp]
    \centering
    {\includegraphics[width=0.8\textwidth]{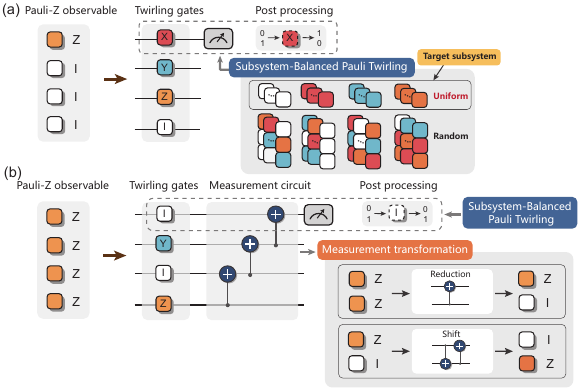}} 
    \caption{{\textbf{The framework for model-free measurement error mitigation (MF-MEM) with subsystem-balanced Pauli twirling (SB-PT).} (a) For sparse Pauli observables, we directly apply SB-PT by uniformly sampling Pauli operators within the observable's support and random Pauli gates on remaining qubits. The first sequence of twirling gates is physically applied; the second is virtually implemented via post process (measurement bit-flips).
    (b) For high-weight observables, we employ measurement transformation to transform the observable into a sparse Pauli form. Then we apply SB-PT to simultaneously twirl the composite noise originating from both the measurement circuit and the native measurement process. The measurement transformation technology comprises two modalities: weight reduction (converting contiguous Pauli-$Z$ operators to sparse ones) and location shift (relocating Pauli-$Z$ operators within the observable).}} 
    \label{fig1}
\end{figure*}

We first introduce the \textbf{Pauli}-$Z$ \textbf{subset}, which consists of index pairs where each pair corresponds to an operator restricted to $\{I, Z\}^{\otimes n}$. Formally, the Pauli-$Z$ subset of index pairs is defined as 
\begin{equation}
    \mathbb{I}_Z := \left\{ (\phi(r),\phi(s)) \mid \vec{r},\vec{s} \in \mathbb{Z}_2^n \right\}.
\end{equation}
For the PTM element $[\mathcal{R}]_{ij}$ with $i,j\in\{0,1,...,4^n-1\}$, it holds that $[\mathcal{R}]_{ij}=0$ when $(i,j)\notin \mathbb{I}_Z$. 
Then we can define the reduced matrix $\mathcal{R}_Z$ as the restriction of the full PTM $\mathcal{R}$ to this Pauli-$Z$ subset $\mathbb{I}_Z$, as illustrated in Fig.~\ref{fig0}(a), which forms a $2^n \times 2^n$ matrix that completely characterizes the classical measurement noise. For $\vec{r},\vec{s}\in\mathbb{Z}_2^n$, the PTM encodes noise statistics through 
\begin{equation}
    [\mathcal{R}]_{\phi(r),\phi(s)}= [\mathcal{R}_Z]_{r,s}=\frac{1}{2^n}\sum_{j=0}^{2^n-1}(-1)^{\vec{s}\cdot \vec{j}}\sum_{k=0}^{2^n-1}[\Lambda]_{kj}(-1)^{\vec{r}\cdot \vec{k}},
\end{equation}
where the dot product $\vec{r}\cdot\vec{k} = \bigoplus_i \vec{r}_i \vec{k}_i$ is computed modulo 2. 
This expression decomposes physically into measurement-induced dephasing channel $\mathcal{D}$ and classical bit-flip noise channel $\mathcal{E}_\Lambda$, as introduced in Eq.~\ref{eq-lambda} and the accompanying text. The term $\sum_j (-1)^{\vec{s}\cdot \vec{j}}$ captures $\mathcal{D}$, representing wavefunction collapse to computational basis states during measurement. Meanwhile, $\sum_k [\Lambda]_{kj}(-1)^{\vec{r}\cdot \vec{k}}$ corresponds to $\mathcal{E}_\Lambda$, computing the expectation value of $Z_r$ after classical readout errors governed by the transfer matrix $\Lambda$.

Especially, under the TPN model, spatial independence simplifies the noise structure significantly. The overall PTM factorizes into single-qubit components:
\begin{equation}
\mathcal{R}_Z = \bigotimes_{k=1}^n \mathcal{R}_Z^{(k)}, \quad \text{where} \quad 
\mathcal{R}_Z^{(k)} = \begin{pmatrix} 1 & 0 \\ \omega_k & \zeta_k \end{pmatrix},
\end{equation}
with local parameters $\omega_k = a_k + b_k - 1$ and $\zeta_k = a_k - b_k$, where $a_k = \mathbb{P}(\text{noisy }0|\text{ideal }0)$ is the probability of obtaining the noisy outcome 0 given the ideal state $|0\rangle$ and $b_k = \mathbb{P}(\text{noisy }1|\text{ideal }1)$ is the probability of obtaining 1 given the ideal $|1\rangle$.
The constraint $|\omega_k| \leq 1-|\zeta_k|$ ensures physical realizability.

This tensor decomposition imposes a strict \textit{trigger condition} that determines which matrix elements can be non-zero. To characterize this condition, we define the support of a Pauli operator $Z_r$ as 
\begin{equation}
    \text{supp}(Z_r) = \{ k \mid \vec{r}_k = 1 \}. 
\end{equation} 
Then, an element $[\mathcal{R}_Z]_{rs}$ of $\mathcal{R}_Z$ is non-zero if and only if the support of $Z_s$ is contained within that of $Z_r$:
\begin{equation}
[\mathcal{R}_Z]_{rs} \neq 0 \iff \text{supp}(Z_s) \subseteq \text{supp}(Z_r).
\end{equation}
Physically, this condition implies that the measurement outcome of $Z_s$ can be affected by $Z_r$ only if $Z_r$ is measured on all qubits included in the measurement of $Z_s$.

The collection of all operator pairs satisfying this support condition defines the \textbf{TPN subset}:
\begin{equation}
\mathbb{T} = \{ (r,s) \mid \text{supp}(Z_s) \subseteq \text{supp}(Z_r) \},
\end{equation}
as illustrated in Fig.~\ref{fig0}(b).

For any $(r,s)\in\mathbb{T}$ and $r\neq 0$, the matrix elements decompose into physically interpretable factors:
\begin{equation}
[\mathcal{R}_Z]_{rs} = \left(\prod_{k:\vec{s}_k=1}\zeta_k\right) \times \left(\prod_{k:\vec{r}_k=1,\vec{s}_k=0}\omega_k\right).
\end{equation}
Here, the first product quantifies bit-flip suppression on qubits measured by both operators, while the second captures the noise contribution from qubits measured by $Z_r$ but not by $Z_s$ (See Appendix~\ref{sec-TPN} for a detailed derivation.).

\subsection{Subsystem-balanced pauli twirling tailored for TPN subspace} 
Building upon the TPN subset $\mathbb{T}$ defined above, we now introduce a decomposition of the noise PTM $\mathcal{R}_Z$ that separates its elements based on membership in $\mathbb{T}$:
\begin{equation}
\mathcal{R}_Z = \mathcal{R}_Z^{\mathbb{T}} + \mathcal{R}_Z^{\overline{\mathbb{T}}},
\end{equation}
where the two components are defined as follows:
\begin{itemize}
    \item $\mathcal{R}_Z^{\mathbb{T}}$ consists of all elements whose indices $(r, s)$ belong to $\mathbb{T}$, i.e., $[\mathcal{R}_Z^{\mathbb{T}}]_{rs} = [\mathcal{R}_Z]_{rs}$ if $(r,s) \in \mathbb{T}$, and $0$ otherwise.
    \item $\mathcal{R}_Z^{\overline{\mathbb{T}}}$ contains all elements whose indices are in the complement set $\overline{\mathbb{T}}$, i.e., $[\mathcal{R}_Z^{\overline{\mathbb{T}}}]_{rs} = [\mathcal{R}_Z]_{rs}$ if $(r,s) \notin \mathbb{T}$, and $0$ otherwise.
\end{itemize}
This decomposition isolates $\mathcal{R}_Z^{\mathbb{T}}$, which captures all local independent errors conforming to the TPN model, as well as a subset of correlated errors consistent with the TPN support structure. The remaining term, $\mathcal{R}_Z^{\overline{\mathbb{T}}}$, contains only correlated errors not captured by the TPN model and is typically weak in magnitude.
 
To specifically target and eliminate the dominant TPN-related errors captured in $\mathcal{R}_Z^{\mathbb{T}}$, we propose Subsystem-Balanced Pauli Twirling (SB-PT). This method constructs tailored twirling sets that are balanced across measured subsystems, enabling efficient mitigation of TPN-structured errors without requiring prior noise characterization. By exploiting the spatial locality of target observables, SB-PT significantly reduces the resource overhead compared to conventional random twirling. The overall structure of MF-MEM combined with SB-PT is illustrated in Fig.~\ref{fig1}(a).

To realize this tailored twirling strategy, we first characterizes the spatial extent of the target observable. We define the weight of a Pauli-$Z$ operator $Z_r$ as the number of qubits on which it acts non-trivially:
\begin{equation}
\tau(r) \equiv |Z_r| = |\text{supp}(Z_r)| = \text{Hamm}(\vec{r}, \vec{0}),
\end{equation}
where $\text{Hamm}(\vec{r},\vec{0})=\sum_{i=1}^n\vec{r}_i$ denotes the Hamming weight of the binary vector $\vec{r}$. This weight $\tau(r)$ corresponds to the number of qubits measured.

The SB-PT set $\mathcal{S}^\text{sub}$ is constructed via a stratified sampling strategy. We choose the size of the twirling set to $|\mathcal{S}^\text{sub}| = c \cdot 4^{\tau(r)}$, where $c \geq 1$ is an integer. Each twirling operator $P_q \in \mathcal{S}^\text{sub}$ is structured as 
\begin{equation}
P_q =
\left( \bigotimes_{i \in \text{supp}(Z_r)} \sigma_{\vec{q}i} \right)
\otimes
\left( \bigotimes_{i' \notin \text{supp}(Z_r)} \sigma_{\vec{q}_{i'}} \right),
\end{equation}
where the first tensor factor acts on the $\tau(r)$-qubit subsystem where $Z_r$ is supported, and the second acts on the remaining $n - \tau(r)$ qubits.
To ensure balanced twirling over the subsystem, the Pauli terms acting on qubits within $\text{supp}(Z_r)$ are selected to satisfy the following condition for each $\Gamma \in \{I, X, Y, Z\}$: 
\begin{equation}
\frac{ |{ q : \sigma_{\vec{q}_i} = \Gamma }| }{ |\mathcal{S}^\text{sub}| } = \frac{1}{4}.
\end{equation}
This ensures uniform sampling over the four Pauli operators for every qubit in the support. For qubits outside $\text{supp}(Z_r)$, the operators are chosen uniformly at random from $\{I, X, Y, Z\}$.

This protocol achieves targeted error suppression by ensuring all error components in $\mathcal{R}_Z^{\mathbb{T}}$ acting on $\text{supp}(Z_r)$ are exactly canceled, as formalized in Theorem~\ref{th:sbpt} (See Appendix~\ref{sec-proof2} for proof.).
 
\begin{theorem}\label{th:sbpt}
For any Pauli-Z observable $Z_r$, we define the trigger set $\mathcal{J}_r$ as a subset of $\{0,...,2^n-1\} \setminus \{r\}$ where
\begin{equation}
    s\in \mathcal{J}_r \Leftrightarrow (r,s)\in\mathbb{T}.
\end{equation}
Then for any $s\in\mathcal{J}_r$, the scaling factor of SB-PT twirling set $\mathcal{S}^\text{sub}$ satisfy:
\begin{equation}
\alpha_{\phi(r)\phi(s)}(\mathcal{S}^\text{sub}) = 0. 
\end{equation}
This induces complete cancellation of all noise components in $\mathcal{R}_Z^\mathbb{T}$ acting on supp($Z_r$).
\end{theorem}

This targeted annihilation within the TPN subset concentrates resources on suppressing the dominant noise component $\mathcal{R}_Z^{\mathbb{T}}$, substantially improving the sampling efficiency for finite-sized twirling sets.
As a necessary consequence, noise components outside this subspace---specifically, elements in $\mathcal{R}_Z^{\overline{\mathbb{T}}}$ and quantum coherent errors---experience weaker suppression compared to random twirling. Nevertheless, by increasing the integer scaling factor $c$, our method progressively suppresses these residual components, driving all error terms toward zero while maintaining balanced Pauli sampling over the subsystem. 

When integrated into the MF-MEM framework, the SB-PT set $\mathcal{S}^\text{sub}$ yields a tighter bound on the estimation error than the general result in Theorem~\ref{theo:tb-bound}, as we establish below in Theorem~\ref{theo:sbpt-bound} (see Appendix~\ref{sec-proof3} for the full proof).

\begin{theorem}\label{theo:sbpt-bound}
We define the extension set as $\Phi(\mathcal{J}_r)=\{\phi(s)|s\in\mathcal{J}_r\}\subseteq\{0,...,4^n-1\}$. For any $n$-qubit quantum state $\rho$ and Pauli-Z observable $Z_r$, with probability $\geq 1-\delta$ over SB-PT set $\mathcal{S}^\text{sub}$, the estimation error $\epsilon(\rho)=|a(\rho)-\text{Tr}(\rho Z_r)|$ is bounded by:
\begin{equation}\label{eq-sbpt_bound}
\epsilon_{\text{SB-PT}}(\rho) \leq \frac{2\kappa_{\text{SB}}(\mathcal{S}^\text{sub}) \sum_{j\notin\Phi(\mathcal{J}_r)} |[\mathcal{R}]_{\phi(r) j}|}{|[\mathcal{R}]_{\phi(r)\phi(r)}| - \kappa_{\text{SB}}(\mathcal{S}^\text{sub}) \sum_{j\notin\Phi(\mathcal{J}_r)}|[\mathcal{R}]_{\phi(r)j}|}
\end{equation}
where 
\begin{equation}
\begin{aligned}
\kappa_{\text{SB}}(\mathcal{S}^\text{sub}) &= \sqrt{\frac{2}{|\mathcal{S}^\text{sub}|} \left(2m\ln 2 + \ln(2/\delta)\right)} \\
m&=n-\tau(r)=n-|\text{supp}(Z_r)|
\end{aligned}
\end{equation}
under the diagonal dominance condition:
\begin{equation}
|[\mathcal{R}]_{\phi(r)\phi(r)}| > \kappa_{\text{SB}}(\mathcal{S}^\text{sub}) \sum_{j\notin\Phi(\mathcal{J}_r)} |[\mathcal{R}]_{\phi(r)j}|.
\end{equation}
This establishes an error bound that is strictly less than the one presented in Eq.~\ref{eq-old_bound} of Theorem~\ref{theo:tb-bound}.
\end{theorem}
The scaling of our bound with $\tau(r)$ complicates direct comparison across observables of different weights, since the size of the twirling set $|\mathcal{S}^\text{sub}| = c \cdot 4^{\tau(r)}$ itself grows with the observable support. Nevertheless, for any fixed $\tau(r)$ and fixed twirling set size, our bound is strictly tighter than that of random twirling (given in Theorem~\ref{theo:tb-bound}), demonstrating the practical benefit of SB-PT under finite sampling.

For high-weight observables, however, the performance gap between SB-PT and random twirling narrows, particularly when $|\mathcal{S}^\text{sub}| < 4^{\tau(r)}$. To mitigate this limitation, we next introduce measurement transformation techniques that reduce the effective weight of such observables. Although these techniques introduce additional circuit noise, they offer a favorable trade-off that preserves overall error mitigation performance, thereby extending the scope of our method to broader classes of observables.

\section{Measurement transformation}\label{sec4}
We propose measurement transformation (MT), a hardware-efficient protocol that converts arbitrary high-weight Pauli-$Z$ observables into measurement-equivalent low-weight counterparts. The protocol integrates two complementary techniques:
\begin{itemize}
\item Weight reduction (WR) compresses contiguous Pauli-$Z$ operators into sparse ones~\cite{Ding2023quantum,Ding2023Compression}, reducing the number of non-trivial Pauli terms.
\item Location shift (LS) spatially redistributes the support of Pauli-$Z$ operators to facilitate WR.
\end{itemize} 
The resulting measurement circuit consists entirely of nearest-neighbor CX gates, as illustrated in Fig.~\ref{fig1}(b). 

\begin{figure*}[!htbp]
    \centering
    {\includegraphics[width=0.74\textwidth]{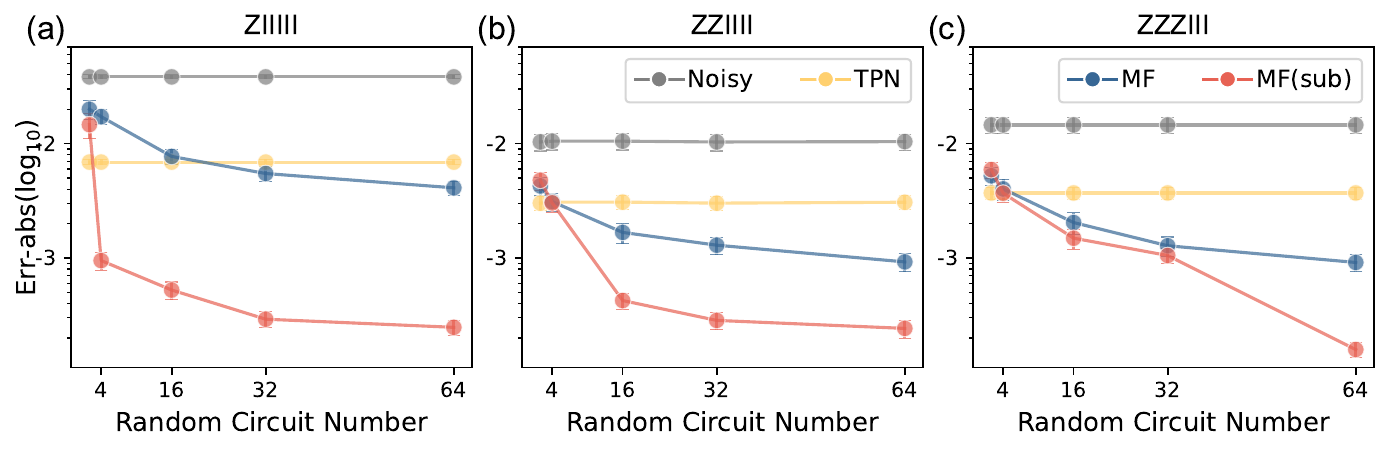}} 
    \caption{{\textbf{The performance comparison between random Pauli twirling and subsystem-balanced Pauli twirling (SB-PT) with increasing random circuit number.} Absolute error between ideal expectation values and estimated results for three six-qubit Pauli observables: (a) $ZIIIII$, (b) $ZZIIII$, and (c) $ZZZIII$, using Haar-random pure states. Four methods are compared: direct noisy measurement (Noisy), measurement error mitigation with tensor product noise model (TPN), model-free mitigation with random Pauli twirling (MF), and model-free mitigation with subsystem-balanced Pauli twirling (MF(sub)). Error bars represent the standard error of the mean (SEM) computed from 100 independent replicates.}}
    \label{fig2}
\end{figure*}
\subsection{Weight reduction}
We begin by structurally decomposing the Pauli-$Z$ observable $Z_r$ into its fundamental components. This involves grouping consecutive non-identity ($Z$) and identity ($I$) operators into contiguous segments.  

Formally, we define the following two sets of segments:
\begin{itemize}
\item ${\Gamma^r_k = Z^{(i_k^1)} \otimes \dots \otimes Z^{(i_k^{\tau_k(r)})}}_k$ is the set of non-identity segments, where each $\Gamma^r_k$ is a tensor product of $\tau_k(r)$ Pauli-$Z$ terms.
\item ${\Upsilon^r_k = I^{(j_k^1)} \otimes \dots \otimes I^{(j_k^{\beta_k(r)})}}_k$ is the set of identity segments, where each $\Upsilon^r_k$ is a tensor product of $\beta_k(r)$ identity terms.
\end{itemize}
The entire observable is then given by the tensor product of all these segments:
\begin{equation}\label{eq-partition}
Z_r = \left(\bigotimes_{k=1}^{n_Z} \Gamma^r_k\right) \otimes \left(\bigotimes_{k=1}^{n_I} \Upsilon^r_k\right),
\end{equation}
The total weight $\tau(r)$ is the sum of the lengths of all non-identity segments: $\tau(r) = \sum_{k=1}^{n_Z} \tau_k(r)$.   

To reduce the measured qubits for the observable $Z_r$, we construct a Clifford circuit $U_R$ that maps $Z_r$ to an effective observable $Z_r^{\text{eff}} = U_R Z_r U^\dagger_R$ with lower weight. This is achieved by compressing each high-weight segment $\Gamma^r_k$ ($\tau_k(r) > 1$) onto a single qubit $t_k$ chosen from its support $\{i_k^1, \dots, i_k^{\tau_k(r)}\}$.

The core component of this compression is a chain of CX gates, denoted as $U_C(i, j)$, which is defined as follows:
\begin{equation}
  U_C(i,j)=\left\{
    \begin{aligned}
        &\text{CX}_{i,i+1}\cdot \text{CX}_{i+1,i+2}\cdot...\cdot\text{CX}_{j-1,j},\ \text{if} \ i<j \\
        &\text{CX}_{(i,i-1)}\cdot\text{CX}_{(i-1,i-2)}\cdot...\cdot \text{CX}_{(j+1,j)},\ \text{if} \ i>j \\
        &I^{\otimes n}, \ \text{if} \ i=j               \\  
    \end{aligned}
  \right. .
\end{equation} 
For each segment $\Gamma^r_k$, we define the Clifford transformation as $U_R(\Gamma^r_k,t_k)=U_C(i_k^{\tau_k(r)},t_k)\cdot U_C(i_k^1,t_k)$, which ensures  
\begin{equation}
U_R(\Gamma^r_k, t_k)\ \cdot \Gamma^r_k\cdot\ U_R^\dagger(\Gamma^r_k, t_k) =  Z^{(t_k)},
\end{equation}
as proven in Appendix~\ref{sec-mc1}. 

The full circuit $U_R$ is then formed by concatenating the segmentwise transformations, yielding the effective observable:
\begin{equation}
Z_r^{\text{eff}} = U_R Z_r U^\dagger_R = \left(\bigotimes_{\tau_k(r)=1} Z^{(i_k^1)}\right) \otimes \left(\bigotimes_{\tau_k(r)>1} Z^{(t_k)}\right) \otimes \left(\bigotimes_{k=1}^{n_I} \Upsilon^r_k \right),
\end{equation}
which has a reduced effective weight $\tau^{\text{eff}} = n_Z$. Measuring $Z_r$ on state $\rho$ is equivalent to measuring $Z_r^{\text{eff}}$ on the state $U_R \rho U_R^\dagger$.

\subsection{Location shift}
However, when the Pauli-$Z$ terms in $Z_r$ are non-contiguously distributed, the direct application of WR is hindered; we therefore introduce LS to address this limitation. It reorganizes the support of $Z_r$ into contiguous segments without altering its total weight, thereby enabling subsequent WR.

The fundamental operation of LS is to shift a $Z$ operator to an adjacent qubit. Consider a pair of adjacent qubits $(i, i+1)$. If the original observable contains $Z^{(i)}$ on qubit $i$ and $I^{(i+1)}$ on qubit $i+1$, we can define the shift operation: 
\begin{equation}
    U_S(i,i+1)=\text{CX}_{i,i+1}\cdot \text{CX}_{i+1,i}.
\end{equation}
The transformation of $Z^{(i)}\otimes  I^{(i+1)}$ under the action of $U_S(i, i+1)$ yields 
\begin{equation}
U_S \cdot\left( Z^{(i)}\otimes  I^{(i+1)} \right) \cdot U_S^\dagger =I^{(i)}\otimes  Z^{(i+1)},
\end{equation}
effectively shifting the $Z$ support from qubit $i$ to qubit $i+1$ (See Appendix~\ref{sec-mc2} for proof.). The reverse shift, from $I^{(i-1)}\otimes Z^{(i)}$ to $Z^{(i-1)}\otimes I^{(i)}$, is achieved by the conjugation operation $U_S(i, i-1)$.

By sequentially applying these nearest-neighbor shift operations, spatially separated Pauli-$Z$ operators can be consolidated into contiguous blocks. Once contiguity is established, the WR protocol can compress each block for full measurement compression.

\subsection{Measurement circuit construction for arbitrary effective Pauli-$Z$ observable}

By integrating the two techniques above, we can construct a Clifford circuit $U_{\text{MT}}$ for any Pauli-$Z$ observable $Z_r$, transforming it into an effective low-weight observable:
\begin{equation}
    Z_r^\text{eff} = U_{\text{MT}} Z_r U^\dagger_{\text{MT}}.
\end{equation}
This allows strategic control over the number and locations of Pauli-$Z$ operators in the measurement basis. Moreover, by redistributing the observable's support, the circuit reduces spatial proximity among measured qubits, thereby inherently suppressing correlated measurement errors that depend on physical layout.
 
The measurement circuit $U_{\text{MT}}$ can be constructed to map $Z_r$ to a single-qubit observable on a target qubit $j \in \{1,\dots,n\}$:
\begin{equation}
U_{\text{MT}} \cdot Z_r\cdot  U^\dagger_{\text{MT}}=I^{\otimes (j-1)} \otimes Z^{(j)} \otimes I^{\otimes (n-j)}.
\end{equation}
This construction is not unique; various sequences of LS and WR operations can achieve the same goal. A typical approach is to first apply LS to group all Pauli-$Z$ terms into one contiguous segment, and then apply WR to compress this segment into a single qubit.

More generally, the framework extends naturally to arbitrary effective weight $\tau^\text{eff} > 1$ by partitioning the qubits into $k$ disjoint subsets $\{H_1, H_2, \dots, H_k\}$, each containing one qubit designated as the measurement site. A similar measurement circuit construction can be applied independently to each subset to achieve the desired low-weight effective observable.

To estimate the expectation value $\langle Z_r\rangle_\rho$, we prepare the state $U_ {\text{MT}}\rho U^\dagger_ {\text{MT}}$, measure in the computational basis, and post-process the outcomes to compute :
\begin{equation}
    \begin{aligned}
        \langle Z_r\rangle_\rho=\text{Tr}(\rho Z_r)=\text{Tr}(\left(U_\text{MT}\rho U_\text{MT}^\dagger\right) Z_r^\text{eff}).
    \end{aligned}
\end{equation}

\subsection{Noise-resilient measurement transformation with Subsystem-Balanced Pauli twirling}

Although MT provides inherent resilience to measurement noise when $\tau^{\text{eff}} < \tau$, the physical implementation of the measurement circuit inevitably introduces additional gate errors. To characterize the effect of noise, we model the measurement circuit under the standard assumption of gate-independent, CPTP noise. Each ideal gate $\mathbb{U}_i$ is accompanied by a noise channel $\mathcal{E}_i$, yielding the noisy implementation:
\begin{equation}
    \tilde{\mathbb{U}}=(\mathcal{E}_l\mathbb{U}_l)\circ...\circ(\mathcal{E}_2\mathbb{U}_2)\circ(\mathcal{E}_1\mathbb{U}_1).
\end{equation}

To analyze the cumulative effect of these errors, we propagate each noise channel to the end of the measurement circuit through conjugation by the subsequent ideal gates. This transformation is achieved as follows: for any noise channel $\mathcal{E}_i$ occurring at the $i$-th step, we conjugate it by all subsequent gates $\mathbb{U}_{i+1:l} = \mathbb{U}_{l} \circ \cdots \circ \mathbb{U}_{i+1}$, effectively shifting it to the end of circuit:
\begin{equation}
    \tilde{\mathbb{U}}=\left(\mathcal{E}_l\circ (\mathbb{U}_{l}\mathcal{E}_{l-1}\mathbb{U}_{l}^\dagger)\circ...\circ(\mathbb{U}_{2:l}\mathcal{E}_1\mathbb{U}_{2:l}^\dagger)\right)\circ\mathbb{U}.
\end{equation}
This expression defines the effective noise channel:
\begin{equation}
    \mathcal{E}^\text{eff}=\mathcal{E}_l\circ (\mathbb{U}_{l}\mathcal{E}_{l-1}\mathbb{U}_{l}^\dagger)\circ...\circ(\mathbb{U}_{2:l}\mathcal{E}_1\mathbb{U}_{2:l}^\dagger),
\end{equation}
which captures the combined effect of all gate errors during the measurement circuit.

In the PTM representation, the noisy measurement circuit is modeled as $\tilde{\mathcal{U}} = \mathcal{C} \cdot \mathcal{U}$, where $\mathcal{C}$ is the PTM of the effective error channel $\mathcal{E}^{\text{eff}}$. The resulting noisy expectation value is given by:
\begin{equation}
    \langle \tilde{Z_r}\rangle_\rho=\langle\langle Z_r^\text{eff}|\mathcal{R}\cdot\mathcal{C}\cdot\mathcal{U}|\rho\rangle\rangle,
\end{equation}
with $\mathcal{R}$ denoting the measurement noise channel. This expression captures the full effect of noise throughout the measurement process and leads to a systematic deviation from the ideal expectation value $\langle Z_r \rangle_\rho$.
The diagonal elements of $\mathcal{C}$, notably $[\mathcal{C}]_{\phi(r)\phi(r)}$, multiply the ideal expectation value by a constant factor. This factor can be readily mitigated through calibration, as shown in Eq.~\ref{eq-tbmem}. However, the off-diagonal terms—including coherent errors—introduce a nontrivial bias that cannot be corrected by simple scaling. 

To address both measurement noise and the gate errors introduced by the measurement circuit, we combine them into a single effective noise channel $\overline{\mathcal{R}} = \mathcal{R} \mathcal{C}$. We then apply SB-PT tailored to the effective observable $Z_r^{\text{eff}}$. For each selected Pauli operator $P_q\in\mathcal{S}^\text{sub}$, we first apply the physically transformed operator $P_{q'}=\mathbb{U}(P_q)=U_\text{MT}P_qU^\dagger_\text{MT}$ prior to the noisy measurement circuit $\tilde{\mathbb{U}}$, and then virtually apply the original $P_q$ during classical post-processing as described in Section~\ref{sec-tbmem}.
This unified mitigation framework suppresses errors in $\overline{\mathcal{R}}$, enabling accurate estimation of $\langle Z_r \rangle_\rho$ even in the presence of coherent errors in both measurement and circuit operations.

\begin{figure*}[t]
    \centering
    {\includegraphics[width=0.79\textwidth]{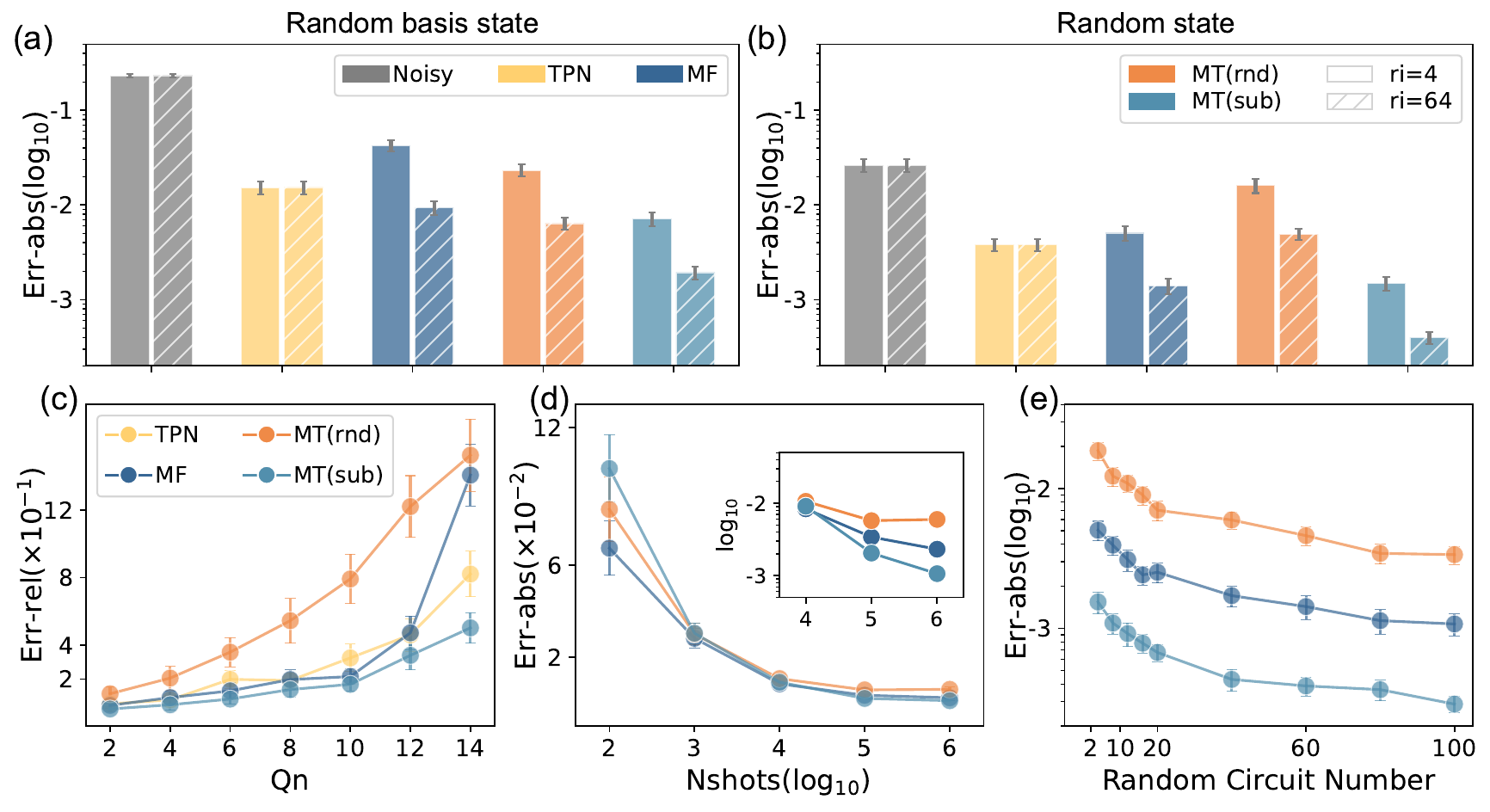}} 
    \caption{{\textbf{Comprehensive performance evaluation of measurement error mitigation methods for global Pauli-Z observables.} Absolute error between ideal expectation values and mitigated estimates is compared across five methods: direct noisy measurement (Noisy), measurement error mitigation with tensor product noise model (TPN), model-free mitigation (MF), and model-free mitigation with measurement transformation (MT) using either random Pauli twirling [MT(rnd)] or subsystem-balanced Pauli twirling [MT(sub)]. (a) Results for six-qubit random computational-basis states with different random circuit numbers (ri = 4 and 64). (b) Results for a six-qubit Haar-random pure states with different random circuit numbers (ri = 4 and 64). (c) Relative error (logarithm of the ratio between mitigated and unmitigated absolute errors) for random states with increasing qubit number (Qn). (d) Performance with increasing measurement shot number for three model-free methods. (e) Performance with increasing random circuit number for three twirling-based model-free mitigation methods: MF, MT(rnd) and MT(sub). For all panels except (d), the measurement shot number is infinite. The measurement noise model is constructed as a tensor product of independent and two-local matrices. Results presented in sub-figure (c) are obtained using synthetic measurement noise model, others are obtained using reconstructed noise from calibrated data on the \textit{Zuchongzhi} 2.1 superconducting quantum processor. And in simulations for sub-figure (c-e), we use Haar-random pure states.   
    Error bars represent the standard error of the mean computed from 100 independent replicates.}} 
    \label{fig3}
\end{figure*}

\section{Numerical Simulations}\label{sec5}

To assess the performance of our method, we perform extensive numerical simulations using two classes of noise-free quantum states: (i) random computational-basis product states, and (ii) Haar-random pure states.

\subsection{Noise Settings}
Measurement noise is systematically incorporated via two distinct types of measurement transfer matrices (See detailed illustration in Appendix~\ref{sec-sim_mem_model}.). The first type is reconstructed from empirical calibration data of qubits $\{Q02, Q05, Q06, Q09, Q10, Q11\}$ on the \textit{Zuchongzhi} 2.1 superconducting quantum processor~\cite{Wu2021Strong,zhu2022Quantum}, which exhibits a measurement error rate (MER) of $0.0452 \pm 0.0143$. 

However, since this experimental characterization is limited to six qubits, we introduce a second, programmable synthetic noise model to investigate larger systems. 
This synthetic noise model incorporates two tunable components, independent and correlated measurement errors, both designed to capture the non-uniform noise profiles observed in real devices. 
The independent part is constructed as a tensor products of single-qubit transfer matrices, where each error strength is independently sampled from a normal distribution with mean $0.015$ and standard deviation $0.01$. Similarly, the correlated component is applied to each of the $n-1$ adjacent qubit pairs via a two-qubit noise operator based on the continuous-time Markov process (CTMP) model~\cite{Bravyi2021Mitigate}:
\begin{equation}
\Lambda=\exp\left(\sum_{i=1}^{4(n-1)}\lambda_iG_i\right),
\end{equation}
where each $\lambda_i$ is also drawn independently from a normal distribution with mean $0.008$ and standard deviation $0.005$. These $\{\lambda_i\}_i$ correspond to the strengths of the generators $\{G_i\}_i$, which belong to four fundamental types of two-qubit readout error processes:
\begin{itemize}
\item $|10\rangle\langle 01|-|01\rangle\langle 01|$ refers to readout error $01\to 10$,
\item $|01\rangle\langle 10|-|10\rangle\langle 10|$ refers to readout error $10\to 01$,
\item $|11\rangle\langle 00|-|00\rangle\langle 00|$ refers to readout error $00\to 11$,
\item $|00\rangle\langle 11|-|11\rangle\langle 11|$ refers to readout error $11\to 00$.
\end{itemize}
The overall synthetic transfer matrix is constructed as a combination of these two noise components. By sampling both independent and correlated error strengths probabilistically, the model introduces qubit-dependent variations that closely emulate the behavior of real superconducting quantum processors.

To model gate noise in the measurement circuit, we employ a hybrid approach combining both incoherent and coherent errors. Incoherent noise is introduced via a global depolarizing channel, with error rates set to $5 \times 10^{-4}$ per single-qubit gate and $5 \times 10^{-3}$ per two-qubit gate. Coherent errors are incorporated using a unitary perturbation of the form $\exp(-i \frac{\beta}{2} XX)$, where the noise strength is set to $\beta = 0.01$.

Unless otherwise stated, the first type of transfer matrix is used by default.

\subsection{Results}
We begin by numerically validating the performance of SB-PT with MF-MEM mentioned in Section~\ref{sec-tbmem}, using six-qubit Haar-random pure states. We evaluate three Pauli-$Z$ observables with increasing weights $\tau$: $ZIIIII$ ($\tau=1$), $ZZIIII$ ($\tau=2$), and $ZZZIII$ ($\tau=3$), comparing the absolute error between ideal expectation values and estimates from four methods: direct noisy measurement (Noisy), MEM with TPN model (TPN), MF-MEM with random twirling (MF), and SB-PT [MF(sub)]. Notably, when the size of the twirling set is less than $4^\tau$, SB-PT strategically constructs the set to ensure that Pauli terms acting on the measured subsystem are non-repeating.

As shown in Fig.~\ref{fig2}, SB-PT significantly outperforms random twirling, achieving superior sampling efficiency and estimation accuracy. A key observation is the pronounced performance leap when the number of random circuits exceeds $4^\tau$. This threshold behavior aligns with the theoretical expectation that SB-PT requires significantly fewer samples than random twirling once the subsystem symmetry is adequately exploited. These results conclusively demonstrate the enhanced efficiency of SB-PT, particularly for low-weight Pauli-$Z$ observables, establishing its advantage over conventional mitigation techniques. 

\begin{figure}[!htbp]
    \centering
    {\includegraphics[width=0.32\textwidth]{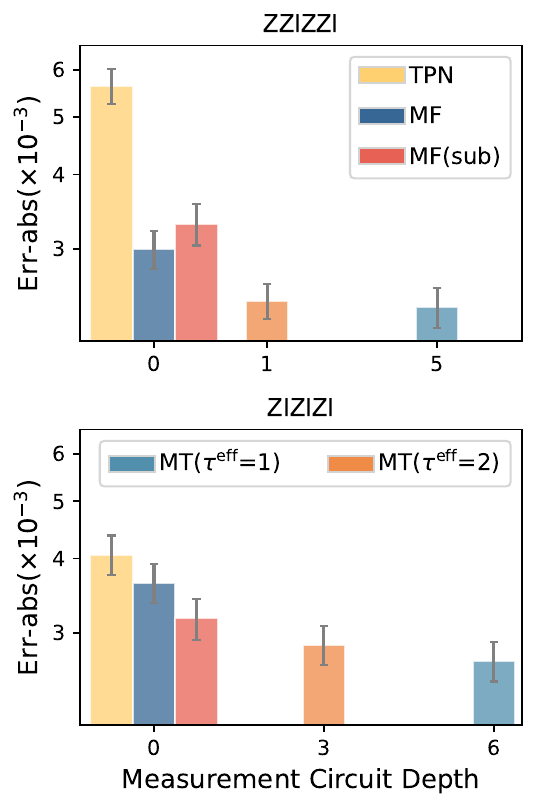}} 
    \caption{{\textbf{Performance comparison of error mitigation methods for structured Pauli-$Z$ observables.} 
    Numerical simulations use ideal six-qubit Haar-random pure states measured under a noise model characterized by calibration data from \textit{Zuchongzhi} 2.1. Two observables are evaluated: $ZZIZZI$ and $ZIZIZI$. MT($\tau^{\text{eff}}=1$) denotes the protocol converting the observable to effective weight-1 via measurement transformation, while MT($\tau^{\text{eff}}=2$) converts to weight-2. Both employ SB-PT. Error bars indicate the standard error of the mean (SEM) over 100 independent replicates.}} 
    \label{fig4}
\end{figure}
We next extend our analysis to global Pauli-$Z$ observables, where the observable's weight equals the system size $n$. In this regime, SB-PT alone fails to provide advantages due to the high weight of the observable. To address this challenge, we incorporate MT to reduce the global observable to weight-1 using a circuit of depth $n-1$ (See Appendix~\ref{sec-circuit} for detailed circuit implementation.). This approach leads to two new hybrid protocols: MF-MEM with MT and random twirling [MT(rnd)], and MF-MEM with MT and SB-PT [MT(sub)]. Notably, the results presented in Fig.~\ref{fig3}(c) are obtained using the second type of synthetic measurement noise model.

As presented in Fig.~\ref{fig3}(a–b), MT(sub) demonstrates a substantial performance advantage, achieving near-optimal error mitigation with very few random circuit instances. Notably, MT(sub) using only ri = 4 attains accuracy comparable to that of the second-best method---MT(rnd) in (a) and MF in (b)---evaluated at ri = 64, corresponding to a 16-fold improvement in sampling efficiency.

The scaling properties of the methods are further investigated using Haar-random states, as shown in Fig.~\ref{fig3}(c-e). Fig.~\ref{fig3}(c) reveals a critical divergence in scaling behavior with increasing qubit number $n$, using the relative error (defined as the logarithm of the ratio between mitigated and unmitigated absolute errors). The relative error for MT(sub) rises gradually, confirming its favorable scaling for larger systems. In contrast, the relative error for the MF method increases abruptly for $n>10$, indicating limited scalability. The shot-number dependence in Fig.~\ref{fig3}(d) demonstrates that the absolute error of MT(sub) continues to improve up to $10^6$ shots, at which point it exceeds the precision of MT(rnd) by an order of magnitude. This confirms its enhanced capability to utilize increased sampling for uncertainty reduction. As shown in Fig.~\ref{fig3}(e), MT(sub) exhibits significantly higher sampling efficiency than all other methods: it achieves at ri = 4 an accuracy comparable to that of MF at ri = 100.
Collectively, these results affirm that the MT(sub) protocol---enabled by SB-PT---offers superior scaling in both system size and sampling efficiency, advancing scalable MEM. Further exploration of the relation between MT(sub) performance and noise parameters is provided in the Appendix~\ref{sec-smresult}.

To examine its flexibility under realistic hardware constraints, where two-qubit gate performance degrades rapidly with circuit depth, we evaluate the trade-off between mitigation accuracy and the depth of measurement circuit using Haar-random states. 

We consider two structured Pauli-$Z$ observables: $o_1 = ZZIZZI$ and $o_2 = ZIZIZI$. For each observable, we compare mitigation using MT with the effective weight reduced to $\tau^\text{eff} = 1$ [MT($\tau^\text{eff} = 1$)] and to $\tau^\text{eff} = 2$ [MT($\tau^\text{eff} = 2$)], both implemented via tailored sequences of WR and LS operations and both employing SB-PT (See more details in Appendix~\ref{sec-circuit}).
For observable $o_1$, the support of the effective observable is reduced to qubit $\{1\}$ under $\tau^\text{eff} = 1$, and to qubits $\{1, 2\}$ under $\tau^\text{eff} = 2$. Similarly, for $o_2$, the support is restricted to qubit $\{1\}$ at $\tau^\text{eff} = 1$, and to qubits $\{1, 3\}$ at $\tau^\text{eff} = 2$.

As shown in Fig.~\ref{fig4}, MT($\tau^\text{eff}$=2) nearly matches the error rates of MT($\tau^\text{eff}$=1), while reducing circuit depth by over 50\%: from 5 to 1 for $o_1$ and from 6 to 3 for $o_2$.
Notably, despite this slight compromise in precision, MT($\tau^\text{eff}=2$) still significantly outperforms conventional methods such as TPN and MF methods. 
These results establish the efficacy of our method in mitigating errors under stringent depth constraints while largely preserving accuracy.

\section{Conclusion}
This work establishes an efficient and scalable framework for measurement error mitigation (MEM) on near-term quantum devices through the introduction of subsystem-balanced Pauli twirling (SB-PT). Designed specifically for model-free measurement error mitigation (MF-MEM), SB-PT leverages the inherent hierarchical structure of measurement noise to achieve high performance for sparse Pauli observables. Its conceptual advance is supported by a rigorously derived sampling bound that is strictly tighter than that of random Pauli twirling under finite sampling.

We further integrate measurement transformation (MT) to extend these advantages to global observables, resulting in a unified architecture that maintains both scalability and hardware efficiency. The resulting framework preserves linear circuit depth in the number of qubits and relies on hardware-friendly nearest-neighbor CX gates, making it especially suitable for superconducting quantum platforms. Crucially, our model offers a unified treatment of native measurement noise and MT-induced gate noise through noise propagation equivalence, enabling holistic error mitigation within a coherent MF-MEM pipeline.

Comprehensive numerical simulations confirm these advances: SB-PT demonstrates a clear performance separation beyond the theoretical threshold of $4^\tau$ random circuits for weight-$\tau$ observables, underscoring its superior sampling efficiency. When integrated with MT, the framework achieves up to a 16-fold reduction in required random circuit instances while preserving accuracy across increasing system sizes. Furthermore, reducing the effective observable weight from 1 to 2 yields over 50\% savings in circuit depth with marginal precision loss, providing practical flexibility for depth-constrained hardware.

Looking forward, this work opens multiple paths for extension—including adaptive MT compilation tuned to specific states and devices, experimental validation on superconducting quantum processors, and generalization to tasks such as one-shot verification. By unifying theoretical soundness with hardware-aware design, SB-PT provides a viable pathway toward scalable quantum advantage in the NISQ era.

\appendix

\section{The proofs of theorems}
\subsection{The proof of Theorem 1}\label{sec-proof1}
For any quantum state $\rho$ and Pauli-$Z$ observable $Z_r$, the estimation error is
\begin{widetext}
\begin{equation}\label{eq-sm_error}
\epsilon(\rho)=|a(\rho)-\text{Tr}(\rho Z_r)|=|\frac{v(\rho)}{v(\rho_0)}-\text{Tr}(\rho Z_r)|
        =\frac{|v(\rho)-\text{Tr}(\rho Z_r)v(\rho_0)|}{|v(\rho_0)|}. 
\end{equation}
\end{widetext}
By expanding the numerator of the given formula, we derive: 
\begin{widetext}
\begin{equation}
\begin{aligned}
    v(\rho)-\text{Tr}(\rho Z_r)v(\rho_0)&=2^n\left( \alpha_{\phi(r)\phi(r)}(\mathcal{S})[\mathcal{R}]_{\phi(r)\phi(r)}\cdot \varrho_{\phi(r)}(\rho)+\sum_{j\neq\phi(r)}\alpha_{\phi(r) j}(\mathcal{S})[\mathcal{R}]_{\phi(r) j}\varrho_j(\rho)-\varrho_{\phi(r)}(\rho)\sum_{j=0}^{4^n-1}\alpha_{\phi(r) j}(\mathcal{S})[\mathcal{R}]_{\phi(r) j})  \right)\\
    &=2^n\left(\sum_{j\neq \phi(r)}\alpha_{\phi(r) j}(\mathcal{S})[\mathcal{R}]_{\phi(r) j}\varrho_j(\rho)-\varrho_{\phi(r)}(\rho)\sum_{j\neq \phi(r)}\alpha_{\phi(r) j}(\mathcal{S})[\mathcal{R}]_{\phi(r) j} \right),
\end{aligned}
\end{equation} 
\end{widetext}
here we use the fact that $\alpha_{\phi(r) \phi(r)}=1$ and $\varrho_j(\rho_0)=1$ for $\rho_0=|0\rangle\langle 0|^{\otimes n}$ and all $j\in\{0,1,...,4^n-1\}$. As $|\varrho_j(\rho)|\leq \frac{1}{2^n}$, we obtain:
\begin{widetext}
\begin{equation}\label{sm-eqt1b1}
\begin{aligned}
    |v(\rho)-\text{Tr}(\rho Z_r)v(\rho_0)|&\leq 2^n\left( \sum_{j\neq\phi(r)}|\alpha_{\phi(r) j}(\mathcal{S})|\cdot |[\mathcal{R}]_{\phi(r) j}|\cdot |\varrho_j(\rho)|+|\varrho_{\phi(r)}(\rho)|\cdot \sum_{j\neq \phi(r)} |\alpha_{\phi(r) j}(\mathcal{S})|\cdot |[\mathcal{R}]_{\phi(r) j}|\right)\\
    &\leq 2\sum_{j\neq \phi(r)}|\alpha_{\phi(r) j}(\mathcal{S})|\cdot |[\mathcal{R}]_{\phi(r) j}| \\
    &\leq 2\max_{j\neq\phi(r)}|\alpha_{\phi(r) j}(\mathcal{S})|\sum_{j\neq\phi(r)}|[\mathcal{R}]_{\phi(r) j}|.
\end{aligned}
\end{equation} 
\end{widetext}

According to the Hoeffding's inequality, for a fixed $j\neq\phi(r)$ and twirling set $\mathcal{S}$ chosen randomly from $\{I,X,Y,Z\}^{\otimes n}$, it gives
\begin{equation}
    \mathbb{P}(|\alpha_{\phi(r) j}(\mathcal{S})|\geq \epsilon)\leq 2\exp\left(-\frac{|\mathcal{S}|\epsilon^2}{2}\right).
\end{equation}
Taking a union bound over $4^n-1$ off-diagonal indices $j\neq\phi(r)$:
\begin{equation}
    \mathbb{P}\left(\max_{j\neq\phi(r)}|\alpha_{\phi(r) j}(\mathcal{S})|\geq \epsilon\right)\leq 2\cdot 4^n\exp\left(-\frac{|\mathcal{S}|\epsilon^2}{2}\right).
\end{equation}
Setting $\delta=2\cdot 4^n\exp\left(-\frac{|\mathcal{S}|\epsilon^2}{2}\right)$ and solving for $\epsilon$:
\begin{equation}
    \epsilon=\sqrt{\frac{2}{|\mathcal{S}|}(2n\ln 2+\ln(2/\delta))}.
\end{equation}
Thus, with probability $\geq 1-\delta$,
\begin{equation}
    \max_{j\neq\phi(r)}|\alpha_{\phi(r) j}(\mathcal{S})|\leq\kappa(\mathcal{S})\stackrel{\Delta}{=}\sqrt{\frac{2}{|\mathcal{S}|}(2n\ln 2+\ln(2/\delta))},
\end{equation}
then we can bound the numerator as
\begin{equation}
    |v(\rho)-\text{Tr}[\rho Z_r]v(\rho_0)|\leq 2\kappa(\mathcal{S})\sum_{j\neq\phi(r)}|[\mathcal{R}]_{\phi(r) j}|.
\end{equation}

Assuming that $[\mathcal{R}]_{\phi(r) \phi(r)}$ is the dominant term (i.e., the diagonal elements of the noise matrix are significantly larger), and the cross terms are relatively small, we derive bounds that are independent of $\rho$. To this end, we assume that the measurement noise $\mathcal{R}$ satisfies $|[\mathcal{R}]_{\phi(r) \phi(r)}|$ being substantially larger than the other terms:
\begin{equation}
    |[\mathcal{R}]_{\phi(r)\phi(r)}|>\kappa(\mathcal{S})\sum_{j\neq \phi(r)}|[\mathcal{R}]_{\phi(r) j}|.
\end{equation}
Then for the denominator of Eq~\ref{eq-sm_error}, we can directly obtain
\begin{equation}\label{sm-eqt1b2}
\begin{aligned}
    |v(\rho_0)|&\geq |[\mathcal{R}]_{\phi(r)\phi(r)}|-\sum_{j\neq\phi(r)}|\alpha_{\phi(r) j}(\mathcal{S})|\cdot|[\mathcal{R}]_{\phi(r) j}|\\
    &\geq |[\mathcal{R}]_{\phi(r)\phi(r)}|-\kappa(\mathcal{S})\sum_{j\neq\phi(r)}|[\mathcal{R}]_{\phi(r)j}|.
\end{aligned}
\end{equation}
Combining the results from Eq.~\ref{sm-eqt1b1} and Eq.~\ref{sm-eqt1b2}, we can derive the upper bound as given in Theorem~\ref{theo:tb-bound}.

\subsection{The proof of Theorem 2}\label{sec-proof2}
For Pauli-$Z$ observable $Z_r$, the scaling factor for any $s\in\mathcal{J}_r$ is
\begin{equation}
    \alpha_{\phi(r)\phi(s)}(\mathcal{S}^\text{sub})=\frac{1}{|\mathcal{S}^\text{sub}|}\sum_{P_q\in \mathcal{S}^\text{sub}}\eta(P_q,Z_rZ_s).
\end{equation}
We denote the restriction of $P_q$ to the subset supp($Z_r$) as $P_q |_{r}=\bigotimes_{i\in\text{supp}(Z_r)}\sigma_{\vec{q}_i}$. Similarly, the restriction to the complement of $Z_r$ is denoted as $P_q|_{\bar{r}}=\bigotimes_{i'\notin \text{supp}(Z_r)}\sigma_{\vec{q}_{i'}}$. 
As for any $s\in\mathcal{J}_r$ supp($Z_s$)$\subseteq$ supp($Z_r$) which indicates $(Z_rZ_s)|_{\bar{r}}=I$, and $(Z_r Z_s)|_r\neq I$ ($Z_r\neq Z_s$), then we can arrive at
\begin{widetext}
\begin{equation}\label{eq-sm_the2}
\begin{aligned} 
    \alpha_{\phi(r)\phi(s)}(\mathcal{S}^\text{sub})&=\frac{1}{|\mathcal{S}^\text{sub}|}\sum_{P_q\in \mathcal{S}^\text{sub}}\eta(P_q |_{r},(Z_r Z_s)|_r)\cdot\eta(P_q |_{\bar{r}},(Z_r Z_s)|_{\bar{r}})\\
    &=\frac{1}{|\mathcal{S}^\text{sub}|}\sum_{P_q\in \mathcal{S}^\text{sub}}\eta(P_q |_{r},(Z_r Z_s)|_r)\cdot\eta(P_q |_{\bar{r}},I)\\
    &=\frac{1}{|\mathcal{S}^\text{sub}|}\sum_{P_q\in \mathcal{S}^\text{sub}}\eta(P_q |_{r},(Z_r Z_s)|_r)\\
    &=0.
\end{aligned}
\end{equation}
\end{widetext}
In the derivation above, the third equality follows from the identity $\eta(P_q, I) = 1$ for all $q \in \{0, \dots, 4^n - 1\}$. The fourth equality is obtained by noting that the restricted operator $P_q |{r} = \bigotimes_{i \in \text{supp}(Z_r)} \sigma_{\vec{q}_i}$ is uniformly distributed over the twirling set $\mathcal{S}^\text{sub}$ of size $c\cdot 4^{\tau(r)}$, thereby forming a complete and balanced sampling of the Pauli subgroup on the support of $Z_r$.

\subsection{The proof of Theorem 3}\label{sec-proof3}
The proof of Eq.~\ref{eq-sbpt_bound} is similar to that in Theorem~\ref{theo:tb-bound}, and we only detail the derivation of $\kappa_\text{SB}(\mathcal{S}^\text{sub})$. As given in Theorem~\ref{th:sbpt}, for any $s\in\mathcal{J}_r$, $\alpha_{\phi(r)\phi(s)}(\mathcal{S}^\text{sub})$ can be regarded as a random variable with
\begin{itemize}
    \item $\mathbb{E}[a_{\phi(r)\phi(s)}]=0$ (for $j\neq\phi(r)$),
    \item $|a_{\phi(r)\phi(s)}|\leq 1$.
\end{itemize}
We can expand the indices in $\mathcal{J}_r$ which satisfies Eq.~\ref{eq-sm_the2} beyond the Pauli-$Z$ subset, since $(Z_r P_q)|_{\bar{r}}=I$ holds for any $\text{supp}(P_q)\subseteq\text{supp}(Z_r)$. We define the extension set as $\Phi(\mathcal{J}_r)=\{\phi(s)|s\in\mathcal{J}_r\}\subseteq\{0,...,4^n-1\}$.
Applying Hoeffding inequality combined with the union bound over $|\overline{\Phi(\mathcal{J}_r)}|=4^{n-\tau(r)}$,
\begin{equation}
    \mathbb{P}\left(\max_{j\notin\Phi(\mathcal{J}_r)}|\alpha_{\phi(s) j}|\geq\epsilon\right)\leq 2\cdot 4^{n-\tau(r)}\exp\left(-\frac{|\mathcal{S}^\text{sub}|\epsilon^2}{2}\right),
\end{equation}
and we can solve for $\epsilon$ via setting 
\begin{equation}
    \delta=2\cdot 4^{n-\tau(r)}\exp\left(-\frac{|\mathcal{S}^\text{sub}|\epsilon^2}{2}\right),
\end{equation} 
we have
\begin{equation}
    \kappa_\text{SB}(\mathcal{S}^\text{sub})\stackrel{\Delta}{=}\epsilon=\sqrt{\frac{2}{|\mathcal{S}^\text{sub}|}(2m\ln 2+\ln(2/\delta))},
\end{equation}
where $m=n-\tau(r)=n-|\text{supp}(Z_r)|$.

We next demonstrate that the error bound in Theorem~\ref{theo:sbpt-bound} is tighter than that in Theorem~\ref{theo:tb-bound}. Defining the function $f(x)=\frac{2x}{a-x}$ for some constant $a$, it is straightforward to verify that $f$ is monotonically increasing for $x<a$. Since $\kappa(\mathcal{S})>\kappa_\text{SB}(\mathcal{S}^\text{sub})$ and $\sum_{j\neq\phi(r)}|[\mathcal{R}]_{\phi(r) j}|\geq\sum_{j\notin\Phi(\mathcal{J}_r)}|[\mathcal{R}]_{\phi(r)j}|$ with twirling set of the same size $|\mathcal{S}|=|\mathcal{S}^\text{sub}|$, we arrive at the conclusion that Theorem~\ref{theo:sbpt-bound} provide a tighter error bound.

\section{The Pauli transfer matrix of the tensor product noise model}\label{sec-TPN}
In this section, we will provide the detailed derivation of the Pauli Transfer Matrix (PTM) structure for the Tensor Product Noise (TPN) model of classical measurement noise.

\subsection{General form of the measurement noise in the PTM representation}
We begin by deriving the general expression for a classical measurement noise $\mathcal{R}$ in the PTM representation. A classical noise model implies that the noisy measurement process can be described by a conditional probability distribution $[\Lambda]_{kj} = \mathbb{P}(\text{noisy} \ j | \text{ideal }k)$, for $k,j\in\{0,...,2^n-1\}$.
The action of $\mathcal{R}$ on an ideal computational basis state $|k\rangle\langle k|$ is
\begin{equation}
    \mathcal{R}|k\rangle\rangle=\sum_{j=0}^{2^n-1}\Lambda_{kj}|j\rangle\rangle,
\end{equation}
where $|k\rangle\rangle$ ($|j\rangle\rangle$) is the PTM representation of state $|k\rangle\langle k|$ ($|j\rangle\langle j|$).
For $p,q\in\{0,...,4^n-1\}$, the element $[\mathcal{R}]_{pq}$, which describes how the noise channel maps the Pauli operator $P_p$ to the component of $P_q$ in the output, is defined as: 
\begin{equation}
\begin{aligned}
    [\mathcal{R}]_{pq}&=\frac{1}{2^n}\langle\langle \mathcal{P}_p|\mathcal{R}|\mathcal{P}_q\rangle\rangle.
\end{aligned}
\end{equation} 
Since the output of $\mathcal{R}$ is always a diagonal matrix in the computational basis ($\sum_j \cdots |j\rangle\langle j|$), the inner product $\langle\langle \mathcal{P}_p|\cdot\rangle\rangle $ will be non-zero only if $P_p$ itself is diagonal in the computational basis. This restricts $P_p$ to the Pauli-$Z$ operators, i.e., $P_p \in \{I, Z\}^{\otimes n}$. This justifies the definition of the Pauli-$Z$ subset $\mathbb{I}_Z$ and the reduced matrix $\mathcal{R}_Z$. 
For the matrix elements $[\mathcal{R}]_{\phi(r),\phi(s)}$ where $P_{\phi(r)}=Z_r$ and $P_{\phi(s)}=Z_s$ are both Pauli-$Z$ operators ($r,s\in\{0,...,2^n-1\}$), we derive:
\begin{widetext}
\begin{equation}
\begin{aligned}
[\mathcal{R}]_{\phi(r),\phi(s)}=[\mathcal{R}_Z]_{rs} &= \frac{1}{2^n} \langle\langle\mathcal{Z}_r|\mathcal{R}|\mathcal{Z}_s\rangle\rangle \\
&= \frac{1}{2^n} \langle\langle\mathcal{Z}_r|\mathcal{R}|\left( \sum_{k=0}^{2^n-1} (-1)^{\vec{s} \cdot \vec{k}} |k\rangle\rangle \right) \\
&= \frac{1}{2^n} \langle\langle\mathcal{Z}_r|\left(\sum_{k=0}^{2^n-1}(-1)^{\vec{s}\cdot\vec{k}}\mathcal{R}|k\rangle\rangle \right)\\
&= \frac{1}{2^n} \langle\langle \mathcal{Z}_r|\left(\sum_{k=0}^{2^n-1}(-1)^{\vec{s}\cdot\vec{k}}\sum_{j=0}^{2^n-1} [\Lambda]_{kj} |j\rangle\rangle \right) \\
&= \frac{1}{2^n} \sum_{k,j} [\Lambda]_{kj} (-1)^{\vec{s} \cdot \vec{k}} \langle\langle\mathcal{Z}_r|j\rangle\rangle \\
&= \frac{1}{2^n} \sum_{k,j} [\Lambda]_{kj} (-1)^{\vec{s} \cdot \vec{k}} (-1)^{\vec{r} \cdot \vec{j}} \quad \text{(since } \langle\langle\mathcal{Z}_r|j\rangle\rangle = (-1)^{\vec{r} \cdot \vec{j}} \text{)} \\
&= \frac{1}{2^n} \sum_{j=0}^{2^n-1} (-1)^{\vec{r} \cdot \vec{j}} \sum_{k=0}^{2^n-1} [\Lambda]_{kj} (-1)^{\vec{s} \cdot \vec{k}},
\end{aligned}
\end{equation}
\end{widetext}
where $\mathcal{Z}_r$ denotes the PTM representation of Pauli-$Z$ operator $Z_r$.

\subsection{Factorization under the TPN Model}
The TPN model assumes the noise is local and uncorrelated across qubits. The conditional probability matrix factorizes as $\Lambda = \bigotimes_{i=1}^n \Lambda^{(i)}$, where the single-qubit response matrix for the $i$-th qubit is: 
\begin{align*}
\Lambda^{(i)}=\begin{pmatrix}
a_i & 1-b_i \\
1-a_i & b_i
\end{pmatrix}=\begin{pmatrix}
\mathbb{P}(0|0) & \mathbb{P}(0|1) \\
\mathbb{P}(1|0) & \mathbb{P}(1|1)
\end{pmatrix}
\end{align*}.

\begin{figure*}[!htbp]
    \centering
    {\includegraphics[width=0.71\textwidth]{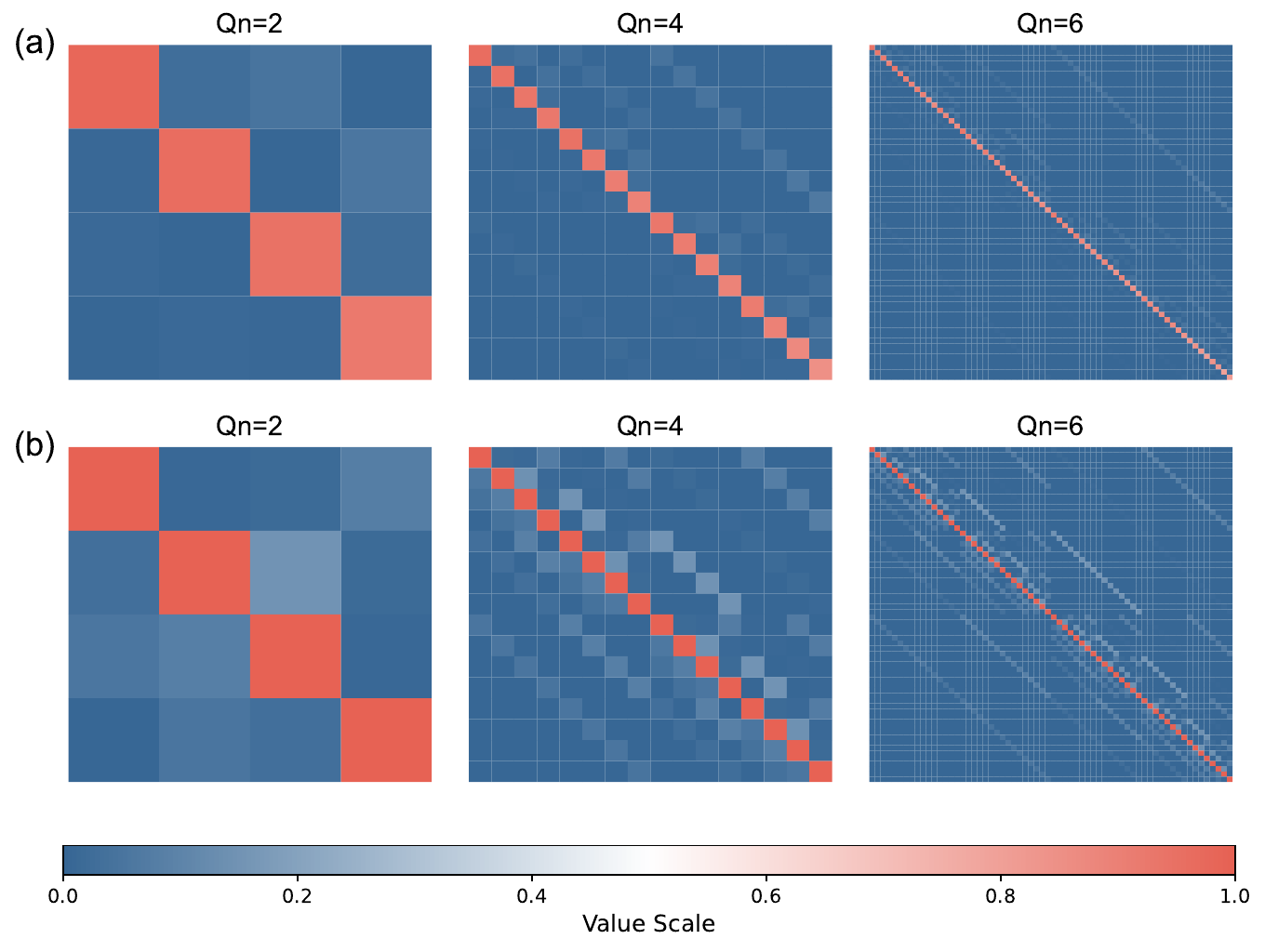}} 
    \caption{{\textbf{The demonstration of measurement noise transfer matrices for varying qubit number.}  
    (a) The transfer matrix reconstructed from empirical calibration data of qubits ${Q02, Q05, Q06, Q09, Q10, Q11}$ on the \textit{Zuchongzhi} 2.1 superconducting processor~\cite{Wu2021Strong,zhu2022Quantum}, which exhibits a mean measurement error rate (MER) of $0.0452 \pm 0.0143$. (b) The programmable synthetic transfer matrix which incorporates dual noise components—independent and correlated measurement errors.}} 
    \label{sm-fig1}
\end{figure*}

We introduce the parameters $\omega_i$ and $\zeta_i$ as defined in the main text:
\begin{align*}
\omega_i &= a_i + b_i - 1, \\
\zeta_i &= a_i - b_i.
\end{align*}
These parameters define the single-qubit PTM $\mathcal{R}_Z^{(i)}$ for the Pauli-$Z$ operators $\{I,Z\}$ on the $i$-th qubit:
\begin{equation}
\mathcal{R}_Z^{(i)} = \begin{pmatrix}
1 & 0 \\
\omega_i & \zeta_i
\end{pmatrix}.
\end{equation} 
Due to the tensor product structure of the noise, the overall reduced PTM $\mathcal{R}_Z$ is the tensor product of the single-qubit PTMs:
\begin{equation}
\mathcal{R}_Z = \bigotimes_{i=1}^n \mathcal{R}_Z^{(i)} = \bigotimes_{i=1}^n \begin{pmatrix} 1 & 0 \\ \omega_i & \zeta_i \end{pmatrix}.
\end{equation}

For the Pauli-$Z$ operators $Z_r=\bigotimes_{i=1}^n Z^{\vec{r}_i}$ and $Z_s= \bigotimes_{i=1}^n Z^{\vec{s}_i}$ ($\vec{r},\vec{s}\in\mathbb{Z}_2^n$), where we denote $Z^0=I$ and $Z^1=Z$, the matrix element $[\mathcal{R}_Z]_{rs}$ is a product over the qubits $i=1$ to $n$:
\begin{equation}
[\mathcal{R}_Z]_{rs} = \prod_{i=1}^n [\mathcal{R}_Z^{(i)}]{\vec{r}_i\vec{s}_i}.
\end{equation} 
For the overall product to be non-zero, we must avoid any factor where $\vec{r}_i=0$ and $\vec{s}_i=1$, as this gives $[\mathcal{R}_Z^{(i)}]_{0,1}=0$. This is precisely the condition $\text{supp}(Z_s) \subseteq \text{supp}(Z_r)$, defining the TPN subset $\mathbb{T}$.
For $(r,s) \in \mathbb{T}$, the product simplifies. We separate the qubits into three groups:
\begin{itemize}
    \item Qubits $i$ where $\vec{s}_i = 1$. Since $(r,s) \in \mathbb{T}$, we must have $\vec{r}_i = 1$ for these qubits. Their contribution to the product is $\prod_{i: \vec{s}_i=1} \zeta_i$.
    \item Qubits $i$ where $\vec{r}_i = 1$ but $\vec{s}_i = 0$. Their contribution is $\prod_{i: \vec{r}_i=1, \vec{s}_i=0} \omega_i$.
    \item Qubits $i$ where $\vec{r}_i = 0$ (and consequently $\vec{s}_i = 0$). Their contribution is $\prod_{i: \vec{r}_i=0} 1=1$.
\end{itemize} 
Thus, the final expression for the matrix element is:
\begin{equation}
[\mathcal{R}_Z]_{rs} = \left( \prod_{i: \vec{s}_i=1} \zeta_i \right) \times \left( \prod_{i: \vec{r}_i=1, \vec{s}_i=0} \omega_i \right), \quad \text{for } (r,s) \in \mathbb{T}.
\end{equation}
This completes the derivation. The first product quantifies the faithful transmission of a $Z$ error on the qubits in the support of $Z_s$, while the second product captures the noise introduced on qubits that are measured but did not have an underlying error.
\begin{figure*}[!htbp]
    \centering
    {\includegraphics[width=0.9\textwidth]{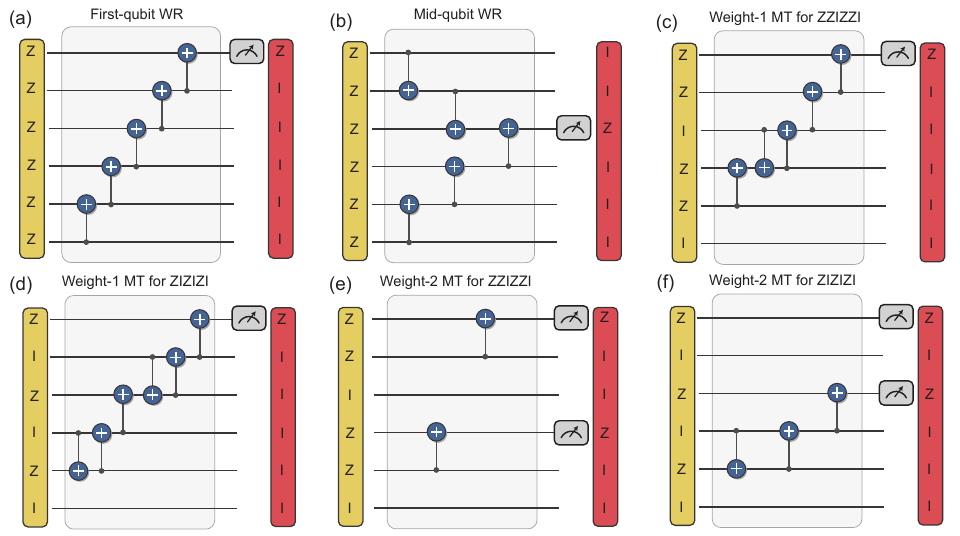}} 
    \caption{{\textbf{The measurement circuits used for measurement transformation.}  
     }Direct measurements of the targeted Pauli-$Z$ observables (yellow) are equivalent to measuring the effective observables (red) via circuits a–f. (a) The circuit for weight reduction (WR) to the first qubit, which transfers global Pauli-$Z$ to weight-1 effective Pauli-$Z$ observable. (b) The circuit for WR to the third qubit. (c) The circuit for weight-1 measurement transformation (MT) for $ZZIZZI$, which transfers to the effective Pauli-$Z$ with support $\{1\}$.  (d) The circuit for weight-1 MT for $ZIZIZI$, which transfers to the effective Pauli-$Z$ with support $\{1\}$. (c) The circuit for weight-2 MT for $ZZIZZI$, which transfers to the effective Pauli-$Z$ with support $\{1,4\}$.  (d) The circuit for weight-2 MT for $ZIZIZI$, which transfers to the effective Pauli-$Z$ with support $\{1,3\}$.} 
    \label{sm-fig2}
\end{figure*}

\section{The measurement circuit used for measurement transformation}
\subsection{The circuit for weight reduction}\label{sec-mc1}
Let $\Gamma^r_k=Z^{(i_k^1)}\otimes...\otimes Z^{(i_k^{\tau_k(r)})}$ be a Pauli-$Z$ operator acting on $\tau_k(r)$ contiguous qubits. The wight reduction (WR) circuit is defined as:
\begin{equation}
  U_R(\Gamma^r_k,t_k)=U_C(i_k^{\tau_k(r)},t_k)\cdot U_C(i_k^1,t_k) ,t_k \in \{i_k^1,...,i_k^{\tau_k(r)}\},
\end{equation}
where $U_C(i,j)$ is a chain of CX gates: 
\begin{equation}
  U_C(i,j)=\left\{
    \begin{aligned}
        &\text{CX}_{i,i+1}\cdot \text{CX}_{i+1,i+2}\cdot...\cdot\text{CX}_{j-1,j},\ \text{if} \ i<j \\
        &\text{CX}_{(i,i-1)}\cdot\text{CX}_{(i-1,i-2)}\cdot...\cdot \text{CX}_{(j+1,j)},\ \text{if} \ i>j \\
        &I^{\otimes n}, \ \text{if} \ i=j               \\  
    \end{aligned}
  \right. .
\end{equation} 
We now show that under this transformation, the effective Pauli observable becomes $Z^{(t_k)}$. 
The conjugate action of a CX gate CX$_{i,j}$, with qubit $i$ as control and qubit $j$ as target, on Pauli operators is:
\begin{equation} 
    \text{CX}_{i,j}\cdot( Z^{(i)}\otimes Z^{(j)})\cdot\text{CX}^\dagger_{i,j}=I^{(i)}\otimes Z^{(j)}. 
\end{equation}
Then for the WR circuit, we have
\begin{widetext}
\begin{equation}
\begin{aligned}
    U_R(\Gamma^r_k,t_k)\cdot \Gamma_k^r\cdot  U^\dagger_R(\Gamma^r_k,t_k)&=U_C(i_k^{\tau_k(r)},t_k) \cdot\left(
    U_C(i_k^1,t_k) \cdot \Gamma_k^r \cdot U^\dagger_C(i_k^1,t_k) \right) \cdot U^\dagger_C(i_k^{\tau_k(r)},t_k)\\
    &=U_C(i_k^{\tau_k(r)},t_k) \cdot\left(
    Z^{(t_k)}\otimes...\otimes Z^{(i_k^{\tau_k(r)})} \right)\cdot U^\dagger_C(i_k^{\tau_k(r)},t_k)\\
    &=Z^{(t_k)}.
\end{aligned}
\end{equation}
\end{widetext}

\subsection{The circuit for location shift}\label{sec-mc2}
The location shift (LS) circuit is defined as:
\begin{equation}
U_S(i,i+1) = \text{CX}_{i,i+1} \cdot \text{CX}_{i+1,i}.
\end{equation}
We now demonstrate that this unitary transformation effects the desired Pauli observable shift:
\begin{equation}
\begin{aligned}
    &U_S(i,i+1) \cdot (Z^{(i)} \otimes I^{(i+1)}) \cdot U^\dagger_S(i,i+1) 
    \\
    &= I^{(i)} \otimes Z^{(i+1)}.
\end{aligned}
\end{equation}
The transformation proceeds in two stages. First, conjugation by $\text{CX}_{i+1,i}$ yields:
\begin{equation}
    \text{CX}_{i+1,i} \cdot(Z^{(i)} \otimes I^{(i+1)}) \cdot\text{CX}^\dagger_{i+1,i} = Z^{(i)} \otimes Z^{(i+1)},
\end{equation}
Subsequent conjugation by $\text{CX}_{i,i+1}$ then produces the final result:
\begin{equation}
\begin{aligned}
    \text{CX}_{i,i+1} \cdot(Z^{(i)} \otimes Z^{(i+1)}) \cdot\text{CX}^\dagger_{i,i+1}= I^{(i)} \otimes Z^{(i+1)}.
\end{aligned}
\end{equation} 
The complementary case follows analogously:
\begin{equation}
U_S(i,i-1) (I^{(i-1)} \otimes Z^{(i)}) U^\dagger_S(i,i-1) = Z^{(i-1)} \otimes I^{(i)}.
\end{equation}
\begin{figure}[!htbp]
    \centering
    {\includegraphics[width=0.48\textwidth]{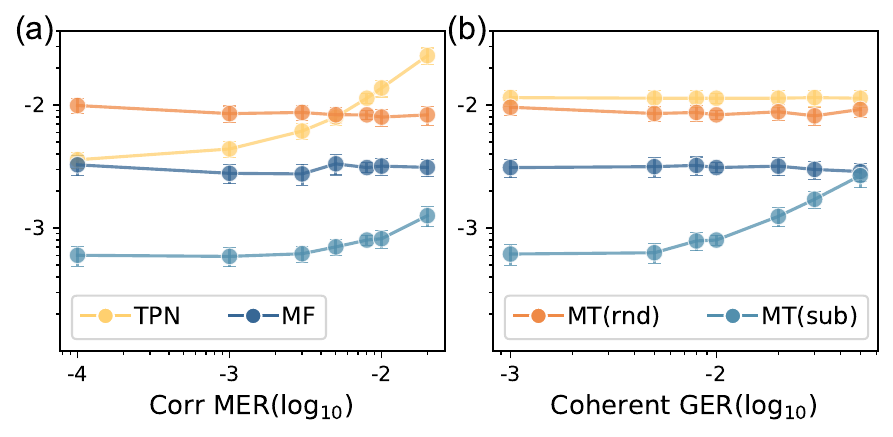}} 
    \caption{{\textbf{Performance of mitigation methods under varying noise parameters for global Pauli-$Z$ observable.} Absolute error is compared across four methods: direct noisy measurement (Noisy), model-free mitigation (MF), and twirling with measurement transformation using either random twirling [MT(rnd)] or subsystem-balanced twirling [MT(sub)]. Simulations use six-qubit noiseless Haar-random states under the synthetic measurement noise model . (a) The absolute error with increasing correlated measurement error rate (Corr MER). (b) The absolute error with increasing coherent gate error rate (Coherent GER). We set the standard Ind MER as 0.015, the Corr MER as 0.008, and the Coherent GER as 0.01.}} 
    \label{sm-fig3}
\end{figure}
\section{Supplementary Simulation Details}
\subsection{The measurement noise model employed in numerical simulations}\label{sec-sim_mem_model}
In our numerical simulations, we employ two distinct classical measurement noise models, both characterized by transfer matrices as illustrated in Fig.~\ref{sm-fig1}.  

\subsection{Circuit implementation for simulations}\label{sec-circuit}
The circuit implementation used in our simulations is illustrated in Fig.~\ref{sm-fig2}. It shows how direct measurements of targeted Pauli-$Z$ observables can be equivalently performed using measurement circuits tailored to their effective observables.

\subsection{Extended simulation results}\label{sec-smresult}
In this section, we investigate the robustness of the MT(sub) method against key noise parameters using a synthetic measurement noise model that incorporates both independent and correlated measurement errors, as well as coherent gate errors in the measurement circuit. We focus specifically on the correlated measurement error rate (MER) and the coherent gate error rate (GER), noting that independent measurement noise and incoherent gate noise are not considered here since they can be effectively suppressed by model-free error mitigation, as discussed in Section~\ref{sec-tbmem}.

As shown in Fig.~\ref{sm-fig3}, MT(sub) exhibits strong resilience to both correlated measurement errors and coherent gate errors at moderate noise levels. Notably, when the correlated MER and GER remain below $10^{-2}$, the performance degradation of MT(sub) is minimal, demonstrating its stability under realistic noise conditions. This robust behavior highlights the practicality of MT(sub) for near-term quantum devices where correlated and coherent errors are non-negligible.

\bibliography{b}
\end{document}